\documentclass[11pt]{article}

\usepackage{amssymb}
\usepackage{amsfonts}
\usepackage{latexsym}

\usepackage{subfigure}
\usepackage{graphics}
\usepackage[dvips,final]{epsfig}
\usepackage{color}

\newcommand{\qa}{\mbox{\quad\mbox{and}\quad}}

\newcommand{\oo}{\mbox{$\mathbb O$}}

\newcommand{\x}{\mbox{$\mathbf x$}}
\newcommand{\y}{\mbox{$\mathbf y$}}

\newcommand{\zz}{\mbox{$\mathbf z$}}

\newcommand{\rt}{\mbox{$\mathbb R$}}

\newcommand{\f}{\mbox{$\mathcal F$}}
\newcommand{\bb}{\mbox{$\mathcal B$}}

\newcommand{\cv}{\mbox{$\mbox{\boldmath $\chi$}$}}
\newcommand{\gm}{\mbox{$\mbox{\boldmath $\gamma$}$}}
\newcommand{\dr}{\mbox{\tiny $\Delta$}}

\newcommand{\ww}{\mbox{$\mathbf w$}}

\newtheorem{remark}{Remark}
\newtheorem{theorem}{Theorem}

\date{}

\begin{document}
 \title{\bf Processing of large sets of stochastic signals: filtering based on piecewise interpolation technique}

 \author{ A. Torokhti  \\
  {\small {\em Centre for Industrial and Applied  Mathematics} }
 \\{\small {\em   University of South  Australia, SA 5095, Australia } }
 \\{\small  anatoli.torokhti@unisa.edu.au} }

\maketitle

\begin{abstract}
Suppose $K_{_Y}$ and $K_{_X}$ are large sets of observed and reference signals, respectively, each containing
$N$ signals.
Is it possible to construct a filter $\f:K_{_Y}\rightarrow K_{_X}$  that requires  a priori information only  on {  few signals}, $p\ll N$, from
$K_{_X}$ but { performs better} than the
known filters based on a priori information on { every} reference signal from  $K_{_X}$? It is shown  that the positive  answer
is achievable under quite unrestrictive assumptions. The device behind the proposed method  is based on a special
extension  of the piecewise linear interpolation technique to the case of random signal sets.
 The proposed technique provides {a single} filter to process any signal from the {arbitrarily large }
signal set.   The filter is determined  in terms of pseudo-inverse matrices so that it always exists.
\end{abstract}
{\bf Keywords:}  Wiener-type  filtering; Interpolation.

\section{Introduction}\label{intr}

A purpose of the proposed   new filtering methodology is to provide an effective way to process large signal sets.
The device behind the proposed method is quite simple and is based on a special
extension  of the piecewise linear interpolation technique to the case of random signal sets. At the same time, such a
device is not straightforward and
requires the  careful substantiation presented in Sections \ref{br-met}, \ref{inter}, \ref{det1} and \ref{error} below.

\subsection{Motivations}\label{mot}

{T}{he} problem under consideration is motivated by the following observations.

\subsubsection{\sc Filtering of large sets of signals; less initial information for  better filtering}\label{mot1}
Suppose we need to transform a set of signals $K_{_Y}$ to another set of signals $K_{_X}$. The signals  are represented by
{ finite} random vectors\footnote{We say a random vector $\x$ is finite if its realization
has a finite number of scalar components.}.
 A major associated difficulty and inconvenience which is common to many known filtering methodologies (see, for example, \cite{che1}-\cite{tor-how},
 \cite{ves1}, \cite{tor-m2}, \cite{rus1,cou1,wig1})
is that they  require a priori information on { each reference signal} to be estimated\footnote{To the best
of our knowledge, the exception is the methodology  in \cite{tor1,tor-m1} where
the filtering  techniques exploit  information on reference signals in the form of the vector obtained from  averaging over
reference signal sets.}.
In particular, the filters in \cite{rus1,cou1,wig1} are based on the use of either the reference signal $\x\in K_{_X}$ itself,
as in  \cite{rus1,cou1}, or its estimate, as in \cite{wig1}.
The Wiener filtering approach (see, for example, \cite{che1}-\cite{tor-m2}, \cite{cou1}, \cite{wig1}) assumes that a covariance matrix formed from a reference signal,
$\x\in K_{_X}$, and an observed signal, $\y\in K_{_Y}$, is known or can be estimated. The latter can be done, for instance, from
samples of $\x$ and $\y$. In particular, this means that the reference signal $\x$ can be measured.

In the case of processing large signal sets, such restrictions become much more inconvenient.

The major motivating question for this work is as follows.
Let $\f:K_{_Y}\rightarrow K_{_X}$ denote a filter  that estimates a large set of reference signals, $K_{_X}$, from  a large
 set of observed signals, $K_{_Y}$. Each set contains $N$ signals.
Is it possible to construct a filter $\f$  that requires  a priori information only  on  {  few signals}, $p\ll N$, from
$K_{_X}$ but { performs better} than the
known filters based on a prior information on { every} reference signal from  $K_{_X}$?
We denote such a filter by $\f^{(p-1)}$.

It is shown in Sections \ref{br-met} and
\ref{error} that the positive  answer is achievable under quite unrestrictive assumptions.  The required features of filter $\f^{(p-1)}$
are satisfied by its special structure described in
Sections \ref{br-met}, \ref{piece} and \ref{inter}. The related conditions are also considered in those Sections.

\subsubsection{\sc  Filtering based on idea of piecewise function interpolation}\label{third}

The specific structure of the proposed filter follows from the extension of piecewise function interpolation
\cite{bab1}.
This is because the technique of piecewise function interpolation \cite{bab1} has significant advantages over the
methods of linear and polynomial approximation  used in  known filtering techniques (such as, for example, those in
\cite{tor100,tor-how}).

The structure of the proposed filter is presented in Sections \ref{br-met}, \ref{piece} and \ref{det1} below.

\subsubsection{\sc Exploiting pseudo-inverse matrices in the filter model}

Most of the  known filtering techniques, for example, those ones in  \cite{che1}-\cite{gol2},  \cite{son1}-\cite{mat1},
\cite{ves1}, \cite{cou1,wig1}, are based on exploiting inverse matrices in their mathematical  models.
In the cases of grossly corrupted signals or erroneous measurements those inverse matrices may not exist and, thus, those
filters cannot be applied. The examples in Section \ref{sim} illustrate this case.

The filter proposed here avoids this drawback since its model is based on exploiting pseudo-inverse matrices.
As a result, the proposed filter always exist. That is, it processes any kind of noisy signals.
An extension of the filtering techniques  to the case of implementation of the  pseudo-inverse matrices is done on the basis of
theory presented in \cite{tor100}.

\subsubsection{\sc Computational work}\label{comp}

Let $m$ and $n$ be the number of components of $\x\in K_{_X}$ and of $\y\in K_{_Y}$, respectively, where $K_{_X}$ and $K_{_Y}$
each contains $N$ signals.
The known filtering techniques  (e.g. see \cite{che1}-\cite{mat1}, \cite{ves1},  \cite{cou1,wig1}), applied to
 $\x$ and $\y$, require the computation of a product of an $m\times n$ matrix and an $n \times n$
matrix, as well as the computation of an $n \times n$  inverse or pseudo-inverse  matrix  for { each pair } of signals
$\x\in K_{_X}$ and $\y\in K_{_Y}$.  This requires   $O(2mn^2)$ and $O(26n^3)$  flops, respectively \cite{gol1}.
 Thus, for the processing of all signals in $K_{_X}$ and $K_{_Y}$, the filters in \cite{che1}-\cite{mat1}, \cite{ves1},
 \cite{cou1,wig1} require $O(2mn^2N)+O(26n^3N)$ operations.

Alternatively,  $K_{_X}$ and $K_{_Y}$  can be represented by vectors, $\cv$ and $\gm$, each with $mN$ and $nN$
components, respectively. In such a case, the techniques in \cite{che1}-\cite{mat1}, \cite{ves1},  \cite{cou1,wig1} can be
applied to $\cv$ and $\gm$ as opposed  to each signals in $K_{_X}$ and $K_{_Y}$. The computational requirement is then
 $O(2mn^2N^2)$ and $O(26n^3N^3)$ operations, respectively \cite{gol1}.

 In both cases, but especially  when $N$ is large, the computational work associated with the approaches
\cite{che1}-\cite{mat1}, \cite{ves1},  \cite{cou1,wig1} becomes  unreasonable hard.

For the filter $\f^{(p-1)}$ to be introduced below, the associated computational work  is substantially less.
 This is because $\f^{(p-1)}$ requires the computation of only $p$ pseudo-inverse matrices associated with  $p$ selected signals in
 $K_{_X}$, where $p$ is much less than the number of signals in
$K_{_X}$. Therefore, for processing of the signal sets,  $K_{_X}$ and $K_{_Y}$,  $\f^{(p-1)}$ requires only
$O(2mn^2p)+O(26n^3p)$ flops where $p\ll N$.
This comparison is illustrated in  Section \ref{sim}.

\subsection{Relevant works}\label{rel}



Some particular filtering techniques  relevant to the  method proposed below are as follows.

\subsubsection{\sc Generic optimal linear (GOL)  filter \cite{tor100}}

The generic optimal linear (GOL)  filter in \cite{tor100} is a generalization of the Wiener filter to the case when
covariance matrix is not invertible and observable signal is arbitrarily noisy (i.e. when, in particular, noise is not
necessarily additive and Gaussian). The GOL filter has been developed for processing an { individual} stochastic signal.
Some ideas from \cite{tor100} are used in the proof of Theorem \ref{th1} below.

\subsubsection{\sc  Simplicial canonical piecewise linear filter \cite{cou1}}

A complex Wiener adaptive filter was developed in \cite{cou1} from the
two-dimensional complex-valued simplicial canonical piecewise linear filter  \cite{jul1}.
The filter in  \cite{cou1} was developed for the  processing of an { individual} stochastic signal   and can be exploited
when the reference  signal is known  and a `covariance-like' matrix is
invertible. The latter precludes an application  to the signal types considered, for example,  in Section \ref{sim}: the matrices used
in  \cite{cou1} are not invertible for the signals as those in Section \ref{sim}.
Similarly, the filters studied in \cite{mat1,ves1} were developed for the processing of
a single signal when the covariance matrices are invertible.

For the filter proposed here, these   restrictions are removed.

\subsubsection{\sc Adaptive piecewise linear filter \cite{rus1}}

A  piecewise linear filter in \cite{rus1} was proposed  for a fixed  image denoising (given by a matrix), corrupted by an
{ additive} Gaussian noise. That is,
the method involved a non stochastic reference signal and required its knowledge.
No theoretical justification for the filter was given in \cite{rus1}.

\subsubsection{\sc Averaging polynomial  filter \cite{tor1,tor-m1}}

The averaging polynomial   filter proposed in \cite{tor1,tor-m1} was developed for the purpose of processing infinite signal sets.
The  filter was based on an argument involving the`averaging' over sets of signals under consideration. This device allows one to
determine a single filter for the processing of  infinite signal sets. At the same time, it leads to an increase in the associated
error when signals differ considerably from each other. This effect is illustrated in Section \ref{sim} below.

\subsubsection{\sc Other relevant filters}\label{other}

The technique developed in   \cite{tor-m2} is an extension of the GOL filter to the constraint problem with respect to the filter
rank. It concerns data compression.


The methods in \cite{son1,che2,kan1,chu1} have been developed for deterministic signals. Motivated by the  results achieved in
\cite{kan1,chu1}, adaptive filters  were elaborated in \cite{lin1}. A theoretical basis for the device proposed in
\cite{kan1,chu1} is provided in \cite{lin2}.

We note that the idea of piecewise linear filtering has been used in the literature in several
{ very different conceptual frameworks},  despite exploiting some very similar terms (as in \cite{kan1}-\cite{jul1}).
At the same time, a common feature of those techniques is that they were developed  for the processing of a single signal,
not of large signal sets as in this paper. In particular, piecewise linear filters in \cite{gel1} have been obtained by
arranging linear filters and thresholds in a tree structure. Piecewise linear filters discussed in \cite{her1} were developed
using so-called threshold decomposition, which is a segmentation operator exploited to split a signal into a set of
multilevel components.
Filter design methods for piecewise linear systems proposed in \cite{fen1} were based on a piecewise Lyapunov function.


\subsection{Difficulties associated with the known filtering techniques}\label{diff11}

Basic difficulties associated with applying the known filtering  techniques to the case under consideration (i.e. to processing of large signal
sets,  $K_{_X}$ and $K_{_Y}$) are that:

(i) they require an information on { each} reference signal (in the form of a sample, for example),

(ii) matrices used in the known filters can be not invertible (as in the simulations considered below in Section \ref{sim})
and then the filter does not exist, and

(ii) the associated  computation work may require  a very long time. For example, in  simulations (Section \ref{sim}),
MATLAB  was out of memory for computing the GOL filter \cite{tor100} when each of sets  $K_{_X}$ and $K_{_Y}$ was represented
by  a long vector (this option has been discussed in Section \ref{comp} above).

\subsection{Differences from  the known filtering techniques}

The differences from the known filtering techniques discussed above are as follows.

(i) We consider a single filter that processes
{ arbitrarily
large} input-output sets of stochastic signal-vectors.
The known filters \cite{che1}-\cite{tor-how},
 \cite{ves1}, \cite{tor-m2}, \cite{kan1}-\cite{wig1} have been developed for the processing of an individual signal-vector only.
In the  case of their application to arbitrarily large signal sets, they imply difficulties described in Sections
\ref{mot} and \ref{diff11} above.

(ii)  As a result, our piecewise linear filter model (Section \ref{filter}), the
statement of the problem (Section \ref{probl} below) and consequently, the device of its solution (Section \ref{main} below) are
different
from those considered in \cite{kan1}-\cite{jul1}. In this regard, see also Section \ref{other}.

(iii) The above naturally leads to a new structure of the filter (presented in Sections \ref{inter} and \ref{det1} below)
which is very  different from the known ones.


\subsection{ Contribution }

In general, for the processing of large data sets, the proposed filter allows us to achieve better results in comparison with the
known techniques in \cite{che1}-\cite{wig1}. In particular, it allows us to

 (i) achieve a desired accuracy in signal estimation\footnote{This means that any desired accuracy is achieved
 theoretically, as is shown in Section \ref{error} below. In practice, of course, the accuracy is increased to
 a prescribed reasonable level.},

 \hspace*{-3mm} (ii) exploit a priori information only  on  { few reference signals}, $p$, from the set
$K_{_X}$ that contains $N\gg p$ signals or even infinite number of signals,

\hspace*{-2mm} (iii) find  { a  single} filter to process any signal from the { arbitrarily large} signal set,

 \hspace*{-3mm} (vi) determine the filter in terms of pseudo-inverse matrices so that the filter always exists, and

 \hspace*{-3mm} (v) decrease the  computational load compared to the related known techniques.


\section{Some preliminaries}\label{some}

\subsection{Notation}\label{not}

The signal sets we consider are, in fact, special representations of time series.

Let $(\Omega, \Sigma, \mu)$ be a probability space\footnote{As usually, $\Omega = \{\omega\}$ is the set of outcomes,
$\Sigma$ a $\sigma$-field of measurable subsets in $\Omega$ and $\mu:\Sigma \rightarrow [0,1]$ an associated
probability measure on $\Sigma$. In particular, $\mu(\Omega) = 1.$},
and   $K_{_X}$ and $K_{_Y}$  be arbitrarily large sets of signals such that
 $$
 K_{_X}=\{ \x(t, \cdot) \in L^{2}(\Omega,{\mathbb R}^{m})\hspace*{1mm}| \hspace*{1mm} t \in T\} \qa
 K_{_Y}=\{ \y(t, \cdot) \in L^{2}(\Omega,{\mathbb R}^{n})\hspace*{1mm}| \hspace*{1mm} t \in T\}
 $$
where $T:=[a,\hspace*{1mm} b]\subseteq \rt.$
We interpret $\x(t, \cdot)$ as a reference signal and $\y(t, \cdot)$ as an observable signal, an input
to the filter $\f$ studied below\footnote{In an intuitive way $\y$ can be
 regarded as a noise-corrupted version of $\x$. For example, $\y$ can be interpreted as $\y = \x +{\mathbf n}$ where
 ${\mathbf n}$ is  white noise. In this paper, we do not restrict ourselves  to this simplest version of $\y$ and assume
 that the dependence of $\y$ on $\x$ and  ${\mathbf n}$ is arbitrary.}.
The variable $t \in T\subseteq \rt$ represents time.\footnote{More generally, $T$ can be
considered as a set of  parameter vectors $\alpha = (\alpha^{(1)}, \ldots, \alpha^{(q)})^T \in C^q\subseteq
\rt^q$, where $C^q$ is a $q$-dimensional cube, i.e., $\y = \y(\alpha, \cdot)$ and $\x = \x(\alpha, \cdot)$.
One coordinate, say $\alpha^{(1)}$ of $\alpha$, could be interpreted as time.}
Then, for example, the random signal $\x(t, \cdot)$ can be interpreted as an arbitrary stationary time series.

Let $\{t_k\}_1^p\subset T$ be a  sequence of fixed time-points such that
\begin{equation}\label{atjb}
a= t_1 < \ldots < t_p = b.
\end{equation}
Because of the partition (\ref{atjb}),
the sets $K_{_Y}$ and $K_{_X}$ are divided in `smaller' subsets $K_{X,1},\ldots, K_{X,{p-1}}$ and $K_{Y,1},\ldots,
K_{Y,{p-1}}$, respectively, so that, for each $j=1,\ldots,p$,
\begin{equation}\label{kxjkyj}
 K_{X,j}=\{ \x(t, \cdot) \hspace*{1mm}| \hspace*{1mm} t_j\leq t \leq t_{j+1}\} \qa
 K_{Y,j}=\{ \y(t, \cdot) \hspace*{1mm}| \hspace*{1mm} t_j\leq t \leq t_{j+1}\}.
\end{equation}
Therefore, $K_{_Y}$ and $K_{_X}$ can now be represented as
\begin{equation}\label{kxky}
K_{_X}= \bigcup _{j=1}^{p-1}K_{X,j}\qa K_{_Y}= \bigcup _{j=1}^{p-1}K_{Y,j}.
\end{equation}

\subsection{Brief description of the problem}

 Given two
{ arbitrarily large} sets of random signals, $K_{_Y}$ and $K_{_X}$, find a single filter
$\f: K_{_Y}\rightarrow K_{_X}$ that estimates the signal $\x \in K_{_X}$ with a controlled, associated
error. Note that in our formulation the set $K_{_Y}$ can be finite or infinite.

\subsection{Brief description of the method}\label{br-met}

The solution of the above problem is based  on the representation of the proposed filter in the form of a sum with  $p-1$ terms
$\f_1,\ldots, \f_{p-1}$  where each term, $\f_j$, is interpreted as  a particular  sub-filter (see  (\ref{fyd}) and (\ref{fydj})
below). Such a filter is denoted by $\f^{(p-1)}:K_{_Y}\rightarrow K_{_X}$.

The sub-filter $\f_j$ transforms signals that belong to `piece'  $K_{Y,j}$ of set $K_{Y}$ to signals in `piece' $K_{X,j}$ of
$K_{X}$, i.e.
$\f_j:K_{Y,j}\rightarrow K_{X,j}$. Each sub-filter $\f_j$  depends on two  parameters, $\alpha_j$ and $\bb_j$.

The prime idea is to determine $\f_j$ (i.e. $\alpha_j$ and $\bb_j$) separately, for each $j=1,\ldots,p-1$. The required
$\alpha_j$ and $\bb_j$ follow from the solutions  of  the equation (\ref{xtp1})
and an associated minimization problem (\ref{xtp1}) (see Sections \ref{inter} and \ref{det1} below).
 This procedure adjusts $\f_j$ so that the  error
associated with the estimation of $\x(t,\cdot)\in K_{X,j}$ is minimal.

A motivation for such a structure of the filter $\f^{(p-1)}$ is as follows. The method of  determining $\alpha_j$ and $\bb_j$
provides an estimate $\f_j[\y(t,\cdot)]$ that
interpolates $\x(t,\cdot)\in K_{X,j}$ at $t=t_j$ and $t=t_{j+1}$. In other words, the filter is flexible to variations in
the sets of observed and reference signals $K_{Y}$ and  $K_{X}$, respectively. Due to this way of determining $\f_j$,
 it is natural to expect that the processing of a `smaller' signal set,  $K_{Y,j}$,   may lead
to a { smaller associated error} than that for the processing
of the  whole  set $K_{Y}$ by a filter which is not specifically adjusted to each particular piece $K_{Y,j}$.

As a result, $\f^{(p-1)}[\y(t,\cdot)]$ represents
a special { piecewise interpolation procedure} and, thus, should be attributed with the associated advantages such as, for example,
 the high accuracy of estimation.

In Section \ref{error}, this observation is confirmed. In Sections \ref{some} and \ref{sim}, it is also shown that the proposed technique allows us to
avoid  the difficulties discussed in Section \ref{diff11} above.

\section{Description of the problem. }\label{filter}

\subsection{Piecewise linear filter model}\label{piece}

Let $\f^{(p-1)}:K_{_Y}\rightarrow K_{_X}$ be a filter such that, for each $t\in T$,
\begin{equation}\label{fyd}
\f^{(p-1)}[\y(t,\cdot)] = \sum_{j=1}^{p-1}\delta_j  \f_j [\y(t,\cdot)],
\end{equation}
 where
 \begin{equation}\label{fydj}
 \f_j[\y(t,\cdot)] = \alpha_j + \bb_j[\y(t,\cdot)]\qa  \delta_j = \left\{ \begin{array}{cl}
                             1,  & \mbox{if $t_j\leq t \leq t_{j+1}$},\\
                             0, & \mbox{otherwise.}
                              \end{array} \right.
\end{equation}
Here, $\f_j$  is a sub-filter defined for $t_j\leq t \leq t_{j+1}$. In (\ref{fydj}),  $\alpha_j= [\alpha_j^{(1)},\ldots , \alpha_j^{(m)}]^T\in  \rt^m$ and $\bb_j: L^2(\Omega,\rt^n) \rightarrow L^2(\Omega,\rt^m)$ is
a linear operator given by a matrix $B_j\in\rt^{m\times n}$, so that
$$
[\bb_j(\y)](t,\omega) = B_j[\y(t,\omega)].
$$
Thus, $\f_j$ is defined by an operator  $F_j: \rt^{n} \rightarrow \rt^{m}$ such that
\begin{equation}\label{fyab}
F_j[\y(t,\omega)] = \alpha_j + B_j[\y(t,\omega)].
\end{equation}

Filter $\f^{(p-1)}$ defined by (\ref{fyd})-(\ref{fyab}) is called the { piecewise } filter\footnote{Hereinafter, we will use
a non-curly symbol to
denote an operator and associated matrix (e.g., the operator $\f_j: L^2(\Omega,\rt^n)
\rightarrow L^2(\Omega,\rt^m)$ and the associated matrix $F_j\in \rt^{m\times n}$ are  denoted by $F_j$).}.

\subsection{Assumptions}\label{assum}

In the known approaches related to filtering of stochastic signals (e.g. see  \cite{che1}-\cite{tor-m2}, \cite{cou1}, \cite{wig1}), it is assumed  that covariance matrices formed from
the reference signal and observed signal are known or can be estimated.

The assumption used here is similar. The covariance matrices that are assumed to be known
or can be estimated, are formed from  { selected} signal pairs $\{\x(t_j,\cdot), \y(t_j,\cdot)\}$ with $j=1,\ldots, p$ and
$p$ to be a small number\footnote{It is worthwhile to note that it is { not} assumed that the covariance matrices are known for
{ each} signal pair from $K_{_X}\times K_{_Y}$, $\{\x(t,\cdot), \y(t,\cdot)\}$ with $t\in [a,\hspace*{1mm} b]$.}, $p\ll N$,
where $N$ is the number of signals
in $K_{_X}$ or $K_{_Y}$.

\subsection{The problem}\label{probl}

In (\ref{fyd})-(\ref{fyab}), parameters of the filter $\f^{(p-1)}$, i.e. vector $\alpha_j$ and matrix $B_j$, for $j=1,\ldots,
p-1$, are unknown. Therefore, under the assumptions described in Section
\ref{assum}, the problem is to determine  $\alpha_j$ and $B_j$, for
$j=1,\ldots, p-1$.
The related problem is to estimate an error associated with the filter $\f^{(p-1)}$.

Solutions to the both problems are given in Sections \ref{det1} and \ref{error}, respectively. In particular,
in the following Section \ref{inter}, interpolation conditions (\ref{xt1}) and (\ref{xtp1})
are introduced that lead to a determination of  $\alpha_j$ and $B_j$.

\subsection{Interpolation conditions}\label{inter}
Let us denote
\begin{equation} \label{nx}
\|\x(t_j,\cdot)\|^2_{\Omega} = {\int}_{\Omega} \|\x(t_j,\omega)\|^2_2 d\mu (\omega)
\end{equation}
where $ \|\x(t_j,\omega)\|_2$ is the Euclidean norm of $\x(t_j,\omega) \in {\mathbb R}^{m}$.

For $t=t_1$, let $\widehat{\x}(t_1,\cdot)$ be an estimate of ${\x}(t_1,\cdot)$ determined by  known methods \cite{che1}-\cite{tor-m2}, \cite{cou1}, \cite{wig1}. This is { the initial condition} of the proposed
technique.

For $j=1,\ldots, p-1$, each sub-filter $F_j$ in  (\ref{fydj})-(\ref{fyab}) is defined so that
$\alpha_j$ and $\bb_j$  satisfy the conditions as follows.

{\em Sub-filter $\f_1$:} For $j=1$,  $\alpha_1$ and  $\bb_{1}$ solve
\begin{eqnarray}\label{xt1}
\widehat{\x}(t_1,\cdot)=\alpha_1 + \bb_1[\y(t_1,\cdot)]\qa
\min_{\small\bb_1} \left \|[\x(t_{2},\cdot)-\alpha_1]  - \bb_1 [\y(t_{2},\cdot)]\right \|^2_\Omega,
\end{eqnarray}
respectively. Then an estimate of ${\x}(t,\cdot)$, $\widehat{\x}(t,\cdot)$, for $t\in [t_1,\hspace*{1mm} t_2]$, is determined as
\begin{eqnarray}\label{xt2ab}
\widehat{\x}(t,\cdot) =  \f_1[\y(t,\cdot)]
=\alpha_1 + \bb_1[\y(t,\cdot)]
=\widehat{\x}(t_1,\cdot) +\bb_1[\y(t,\cdot)-\y(t_1,\cdot)]
\end{eqnarray}
where $\alpha_1$ and $\bb_1$  satisfy (\ref{xt1}). In particular, $\alpha_1=\widehat{\x}(t_1,\cdot)-\bb_1[\y(t_1,\cdot)]$ and
$$\widehat{\x}(t_2,\cdot)=\f_1[\y(t_{2},\cdot)].$$
Extending this procedure up to $j=k-1$, where $k=3,\ldots,p$, we set the following. Let $\widehat{\x}(t_{k-1},\cdot)$ be an
estimate of ${\x}(t_{k-1},\cdot)$ defined by the preceding steps as
\begin{eqnarray}\label{xtp1fp1}
\widehat{\x}(t_{k-1},\cdot)=\f_{k-2}[\y(t_{k-1},\cdot)].
\end{eqnarray}
Then sub-filter $\f_{k-1}$ is defined as follows.

{\em Sub-filter $\f_{k-1}$:} For $j=k-1$, $\alpha_{k-1}$ and  $\bb_{k-1}$ solve
\begin{eqnarray}\label{xtp1}
\widehat{\x}(t_{k-1},\cdot)=\alpha_{k-1} + \bb_{k-1}[\y(t_{k-1},\cdot)]\qa
\min_{\small\bb_{k-1}} \left \|[\x(t_{k},\cdot)-\alpha_{k-1}]  - \bb_{k-1} [\y(t_{k},\cdot)]\right \|^2_\Omega,
\end{eqnarray}
respectively.
Then an estimate of ${\x}(t,\cdot)$, $\widehat{\x}(t,\cdot)$, for $t\in [t_{k-1},\hspace*{1mm} t_k]$, is determined as
\begin{eqnarray}\label{xtjab}
\widehat{\x}(t,\cdot) =  \f_{k-1}[\y(t,\cdot)]
=\alpha_{k-1} + \bb_{k-1}[\y(t,\cdot)]
=\widehat{\x}(t_{k-1},\cdot) +\bb_1[\y(t,\cdot)-\y(t_{k-1},\cdot)].
\end{eqnarray}
The conditions (\ref{xt1}) and (\ref{xtp1}) are motivated by the device of piecewise function
interpolation and associated advantages \cite{bab1}.

Filter $\f^{(p-1)}$ of the form (\ref{fyd})-(\ref{fydj}) with $\alpha_j$ and $\bb_j$ satisfying  (\ref{xt1}) and (\ref{xtp1})
 is called the {\em piecewise linear interpolation} filter.
The pair of signals $\{\x(t_{k},\cdot), \y(t_{k},\cdot)\}$ associated with  time $t_k$ defined by (\ref{atjb}) is called
the {\em interpolation pair}.

\section{Main results} \label{main}

\subsection{General device}\label{general}

In accordance withe the scheme presented in Sections \ref{piece} and \ref{inter} above, an estimate of the reference signal
$\x(t,\cdot)$, for any $t\in T=[a,\hspace*{1mm} b]$,  by the  { piecewise linear interpolation} filter $\f^{(p-1)}$, is given by
\begin{eqnarray}\label{xtfyt}
\widehat{\x}(t,\cdot)=\f^{(p-1)}[\y(t,\cdot)] = \sum_{j=1}^{p-1}\delta_j  \f_j [\y(t,\cdot)],
\end{eqnarray}
where, for each $j=1,\ldots,p-1,$ the sub-filter $\f_j$ is given by (\ref{fydj}), and is defined from the interpolation
conditions  (\ref{xt1}) and (\ref{xtp1}).

Below, we show how to determine  $\f_j$ to satisfy the conditions  (\ref{xt1}) and (\ref{xtp1}).

\subsection{Determination of piecewise linear interpolation filter}\label{det1}


Let us denote
\begin{eqnarray}\label{zjwj}
 && \zz(t_j, t_{j+1},\cdot)= \x(t_{j+1},\cdot)-\widehat{\x}(t_{j},\cdot) \qa \ww(t_j, t_{j+1},\cdot)= \y(t_{j+1},\cdot)-\y(t_{j},\cdot).
\end{eqnarray}
We need to represent $\zz(t_j, t_{j+1},\cdot)$ and $\ww(t_j, t_{j+1},\cdot)$ in terms of their components as follows:
\begin{eqnarray*}
&& \zz(t_j, t_{j+1},\cdot)=[\zz^{(1)}(t_j, t_{j+1},\cdot),\ldots, \zz^{(m)}(t_j, t_{j+1},\cdot)]^T \\
\mbox{and} && \ww(t_j, t_{j+1},\cdot)=[\ww^{(1)}(t_j, t_{j+1},\cdot),\ldots, \ww^{(n)}(t_j, t_{j+1},\cdot)]^T,
\end{eqnarray*}
where $\zz^{(j)}(t_j, t_{j+1}, \cdot) \in L^{2}(\Omega,{\mathbb R})$ and $\ww^{(i)}(t_j, t_{j+1}, \cdot) \in L^{2}(\Omega,{\mathbb R})$
are random variables, for all $j=1,\ldots,m$.

Then we can introduce the covariance matrix
\begin{eqnarray}\label{w3}
E_{z_{j}w_{j}} = \left \{\left\langle \zz^{(i)}(t_j,t_{j+1},\cdot), \ww^{(k)}(t_j,t_{j+1},\cdot)\right\rangle
\right \}_{i,k=1}^{m, n},
\end{eqnarray}
where
$
 \left\langle  \zz^{(i)}(t_j, t_{j+1},\cdot),  \ww^{(k)}(t_j, t_{j+1},\cdot)\right\rangle
= \int_{\Omega}  \zz^{(i)}(t_j,t_{j+1},\omega)  \ww^{(k)}(t_j, t_{j+1},\omega)\ d\mu (\omega).
$

Below, $M^{\dag}$ is the Moor-Penrose generalized inverse of a matrix $M$.

Now, we are  in a position to establish the main results.

\begin{theorem}\label{th1} Let
 $$K_{_X}=\{ \x(t, \cdot) \in L^{2}(\Omega,{\mathbb R}^{m})\hspace*{1mm}| \hspace*{1mm} t \in T=
[a, \hspace*{1mm} b]\}\qa K_{_Y}=\{ \y(t, \cdot) \in L^{2}(\Omega,{\mathbb R}^{n})\hspace*{1mm}| \hspace*{1mm} t \in T=[a,
\hspace*{1mm} b]\}$$  be  sets of reference signals and  observed signals, respectively.
Let  $t_j\in [a, \hspace*{1mm} b]$, for $j=1,\ldots,p$, be such that
$$
a=t_1 < \ldots < t_p = b.
$$
 For $t=t_1,$ let $\widehat{\x}(t_1,\cdot)$ be a known estimate of ${\x}(t_1,\cdot)$\footnote{As it has been mentioned in
 Section \ref{inter}, $\widehat{\x}(t_1,\cdot)$  can be determined by the known methods.}.
 Then, for any $t\in [a, \hspace*{1mm} b]$,  the  proposed piecewise linear interpolation filter $\f^{(p-1)}:L^2(\Omega,\rt^n)
 \rightarrow L^2(\Omega,\rt^m)$ transforming any signal $\y(t,\cdot)\in L^2(\Omega,\rt^m)$ to an estimate of $\x(t,\cdot)$,
 $\widehat{\x}(t,\cdot)$, is given by
\begin{equation}\label{fyd0}
\widehat{\x}(t,\cdot) ={\f^{(p-1)}}[\y(t,\cdot)] = \sum_{j=1}^{p-1}\delta_j  {\f}_j [\y(t,\cdot)]
\end{equation}
where
\begin{eqnarray}\label{f0jxj}
\hspace*{-7mm}{\f}_j [\y(t,\cdot)] =  \widehat{\x}(t_j,\cdot)  + \bb_j[\y(t,\cdot) - \y(t_j,\cdot)],
\end{eqnarray}
\begin{eqnarray}\label{}
\widehat{\x}(t_j,\cdot) = {\f}_{j-1} [\y(t_j,\cdot)]\quad \mbox{(for $j=2,\ldots,p-1$)},
\end{eqnarray}
\begin{eqnarray}\label{b0j}
&&\hspace*{-9mm} B_j = E_{z_{j}w_{j}}E_{w_{j}w_{j}}^{\dag}+ M_{B_j}[I_n- E_{w_{j}w_{j}} E_{w_{j}w_{j}}^{\dag}],
\end{eqnarray}
and  where  $I_n$ is the $n\times n$ identity matrix and $M_{B_j}$ is an $m\times n$ arbitrary matrix.
\end{theorem}

{\em Proof}: The proof of Theorem \ref{th1} is given in Section \ref{appa}. $\hfill \Box$

It is worthwhile to observe that, due to an arbitrary matrix $M_{B_j}$ in (\ref{b0j}),  the filter $\f^{(p-1)}$ is not unique.
In particular, $M_{B_j}$ can be chosen as the zero matrix $\oo$ similarly to the generic optimal linear \cite{tor100} (which is also not
unique by the same reason).

\subsection{Numerical realization of filter  $\f^{(p-1)}$ and associated algorithm }\label{alg}

\subsubsection{Numerical realization }\label{numer}

In practice, the set $T=[a,\hspace*{1mm} b]$ (see Section \ref{not}) is represented by a finite set
$\{\tau_{k}\}_{k=1}^N$, i.e. $[a,\hspace*{1mm} b] = [\tau_{1},\tau_{2},\ldots, \tau_{N}]$ where
$a\leq \tau_{1} < \tau_{2}<\ldots < \tau_{N}\leq b$.

For $k=1,\ldots,N$, the estimate of $\x(\tau_{k},\cdot)$, $\widehat{\x}(\tau_{k},\cdot)$,  and  observed signal $\y(\tau_{k},\cdot)$
are  represented  by $m \times q$ and  $n \times q$ matrices
\begin{eqnarray}\label{}
\widehat{X}^{(k)}=[\widehat{\x}(\tau_{k},\omega_1), \ldots, \widehat{\x}(\tau_{k},\omega_{q})]\hspace{3mm}\mbox{and}\hspace{3mm}
Y^{(k)}=[\y(\tau_{k},\omega_1), \ldots, \y(\tau_{k},\omega_{q})].
\end{eqnarray}
 The sequence of fixed time-points $\{t_k\}_1^p\subset [a,\hspace*{1mm} b]$ introduced   in (\ref{atjb}) is such that
\begin{equation}\label{tatj}
\tau_{1}= t_1 < \ldots < t_p = \tau_{N},
\end{equation}
where
$
t_1=\tau_{n_0}, \quad  t_2=\tau_{n_0+n_1},  \quad  \ldots, \quad  t_p=\tau_{n_0+n_1\ldots+n_{p-1}},
$
 and where $n_0=1$ and $n_1,\ldots, n_{p-1}$ are positive integers  such that $N=n_0+n_1+\ldots +n_{p-1}$.

For $j=1,\ldots, p$, vectors  $\widehat{\x}(tj,\cdot)$ and  $\y(tj,\cdot)$ associated with $t_j$ in (\ref{tatj}) are represented,
respectively,  by
 $$
\widehat{X}^{(k)}_j=[\widehat{\x}(t_j,\omega_1), \ldots, \widehat{\x}(t_j,\omega_{N})]\qa
Y_j=[\y(t_j,\omega_1), \ldots, \y(t_j,\omega_{N})].
$$

\subsubsection{Algorithm }

As it has been mentioned in Section \ref{inter}, it is supposed that, for $t=t_1,$  an estimate of $X_1$,  $\widehat{X}_1$, is
known and can be determined by the known methods. This is the initial condition of the proposed technique.

On the basis of the results obtained in  Sections \ref{inter} and \ref{det1}, the performance algorithm of the proposed filter
 consists of the following steps. For $j=1\ldots,p$, we write $N_j=n_0+n_1+\ldots +n_{j-1}$.
\medbreak

{\em Initial parameters:} $Y^{(1)},\ldots,Y^{(N)}$, $\{t_j\}_{j=1}^p$ (see (\ref{tatj})), $\{E_{z_{j}w_{j}}\}_{j=1}^p$,
$\{E_{w_{j}w_{j}}\}_{j=1}^p$
(see (\ref{zjwj}) and (\ref{w3})),  $\widehat{X}_1$, $n_0=1$ , $N_0=0$ and $M_{B_j}=\oo$, for $j=1,\ldots,p-1$.

(Possible ways to get estimates of $E_{z_{j}w_{j}}$ and  $E_{w_{j}w_{j}}$ are discussed below in Section \ref{some}.)

{\em Final parameters:} $\widehat{X}^{(2)}$, $\widehat{X}^{(3)}$,$\ldots$, $\widehat{X}^{(N)}$.

{\em Algorithm}:

$\bullet$ for $j=1$ to $p$ do

\hspace*{3mm}begin
$$
B_j = E_{z_{j}w_{j}}E_{w_{j}w_{j}}^{\dag};
$$

\hspace*{5mm} $\bullet$ for $k=N_{j-1}+1$ to $N_{j}$ do

\hspace*{9mm}begin
$$
\widehat{X}^{(k)} = \widehat{X}_j + B_j (Y^{(k)} - Y_j);
$$
$$
 \widehat{X}_{j+1} = \widehat{X}^{(N_j)};
$$
\hspace*{9mm}end

\hspace*{3mm}end

\subsection{ Error analysis }\label{error}

It is natural to expect that the error associated with the piecewise interpolating filter $\f^{(p-1)}$
decreases when  $\displaystyle \max_{j=1,\ldots,p-1} \dr t_j$ decreases.
 Below, in Theorem \ref{th3}, we justify that this observation is true.
To this end, first, in the following Theorem \ref{th2}, we establish an estimate of the error associated with the filter $F$.

Let us introduce the norm by
\begin{eqnarray}\label{norm}
\|\x(t,\cdot)\|^2_{T,\Omega} =\frac{1}{b-a}{\int}_T \|\x(t,\cdot)\|^2_{\Omega} dt.
\end{eqnarray}
We also denote $\|\x(t,\omega)\|^2_{T,\Omega}=\|\x(t,\cdot)\|^2_{T,\Omega}$.

 Let us suppose that $\x(\cdot,\omega)$ and $\y(\cdot,\omega)$ are Lipschitz continuous signals, i.e.
that there exist real non-negative constants $\lambda_{j}$ and $\gamma_{j}$, with $j=1,\dots,p $, such that, for
$t\in [t_j,\hspace*{1mm} t_{j+1}]$,
\begin{eqnarray}\label{lip11}
\|\x(t,\omega) -\x(t_j,\omega)\|^2_{{T,\hspace*{1mm} \Omega}}\leq \lambda_{j} \dr t_j \qa
\|\y(t,\omega) -\y(t_{j+1},\omega)\|^2_{{T,\hspace*{1mm} \Omega}}\leq \gamma_{j}\dr t_j
\end{eqnarray}
where $\dr t_j =|t_{j+1} - t_j|$.

 \begin{theorem}\label{th2} Under the conditions (\ref{lip11}) the error associated with the piecewise interpolation filter,
 $\|\x(t,\omega) - {F}^{(p-1)}[\y(t,\omega)]\|^2_{{T,\hspace*{1mm} \Omega}}$, is estimated as follows:
\begin{eqnarray}\label{er13}
&&\hspace*{-15mm}\|\x(t,\omega) - {F}^{(p-1)}[\y(t,\omega)]\|^2_{{T,\hspace*{1mm} \Omega}}\leq  \max_{j=1,\ldots, p-1}\left[(\lambda_{j}
+\gamma_{j}\|B_j\|^2)\dr t_j
 + \|E_{z_{j}z_{j}}^{1/2}\|^2 - \|E_{z_{j}w_{j}}(E_{w_{j}w_{j}}^{1/2})^\dag\|^2\right].
\end{eqnarray}
\end{theorem}
{\em Proof}: The proof of Theorem \ref{th2} is given in Section \ref{appa}. $\hfill \Box$

\medbreak
Further, to show that the error of the reference signal estimate  tends to the zero, we need to assume that,
 for $t\in [t_1, \hspace*{1mm} t_2]$, the known estimate
$\widehat{\x}(t_{1},\omega)$ differs from ${\x}(t,\omega)$ for the value of the order $\dr t_1$, i.e. that, for some
constant $c_1\geq 0$,
\begin{eqnarray}\label{c1}
\| \x(t,\omega) - \widehat{\x}(t_{1},\omega)\|^2_{\Omega} \leq c_1 \dr t_1, \quad \mbox{for $t\in [t_1, \hspace*{1mm} t_2]$}.
\end{eqnarray}

\begin{theorem}\label{th3} Let the conditions $(\ref{lip11})$ and $(\ref{c1})$ be true. Then the error  associated with
the piecewise interpolating filter $F$, $\|\x(t,\omega) - {F}^{(p-1)}[\y(t,\omega)]\|^2_{{T,\hspace*{1mm} \Omega}}$,
decreases in the following sense:
\begin{eqnarray}\label{er17}
\|\x(t,\omega) - {F}^{(p-1)}[\y(t,\omega)]\|^2_{{T,\hspace*{1mm} \Omega}} \rightarrow 0
\quad\mbox{as}\hspace*{2mm} \displaystyle \max_{j=1,\ldots,p-1} \dr t_j\rightarrow 0\qa p  \rightarrow \infty.
\end{eqnarray}
\end{theorem}

\noindent
{\em Proof}: The proof of Theorem \ref{th3} is given in Section \ref{appa}. $\hfill \Box$

\begin{remark}\label{rem2}
We would like to emphasize that the statement of Theorem \ref{th3} is fulfilled  only under assumptions
(\ref{lip11}) and (\ref{c1}).
At the  same time, the assumptions (\ref{lip11}) and (\ref{c1})  are not restrictive from a
practical point of view. The condition (\ref{lip11}) is true for Lipschitz continuous signals $\x$ and $\y$, i.e. for very wide class
of signals. The condition  (\ref{c1}) is achieved by a choosing an appropriate known method  (e.g. see  \cite{che1}-\cite{tor-m2}, \cite{cou1}, \cite{wig1})
to find the estimate $\widehat{\x}(t_{1},\omega)$ used in the proposed filter $\f^{(p-1)}$ (see  (\ref{xt1}) and Theorem \ref{th1}).

\end{remark}

\subsection{Some remarks related to the assumptions of the method}\label{some}

As it has been mentioned in Section \ref{assum},  for $j=1,\ldots, p$, matrices $E_{z_j w_j}$ and $E_{w_j w_j}$ in (\ref{b0j}) are
assumed to be known or can be estimated. Here, $p$ is a chosen number of selected interpolation signal pairs (see Section
\ref{inter}). We note that normally $p$
is much smaller than the number of input-output signals $\x(t,\cdot)$ and $\y(t,\cdot)$. Therefore, to estimate any signal $\x(t,\cdot)$ from an arbitrarily large
set $K_{_X}$, only a small number, $p$, of matrices $E_{z_j w_j}$ and $E_{w_j w_j}$ should be estimated (or be known).
This issue has also been discussed in Sections \ref{mot1} and \ref{comp}.

By the proposed method, $\x(t,\cdot)$ is estimated for $t\in [t_j, \hspace*{1mm} t_{j+1}]$. While $E_{w_j w_j}$ in (\ref{b0j}) can be directly estimated from  observed signals $\y(t_{j+1},\cdot)$ and $\y(t_{j},\cdot)$, an estimate
of matrix $E_{z_j w_j}$ depends on the reference signal $\x(t_{j+1},\cdot)$ (see (\ref{zjwj}) and (\ref{w3})) which is unknown (because
the estimate is considered for $t\in [t_j, \hspace*{1mm} t_{j+1}]$).

Some possible approaches to an estimation of matrix $E_{z_j w_j}$ could be as follows.

1. In the general case, when $\x(t,\cdot)$ and  $\y(t,\cdot)$
are arbitrary signals as discussed in Section \ref{not} above, matrix $E_{z_j w_j}$ can be estimated as proposed, for example,
 in \cite{and1}, from samples of  $z_j$ and $w_j$.

2. In the case of incomplete observations, the method proposed in \cite{per1,led1} can be used.

3. Let $E_{\hat{z}_j w_j}$ be a matrix obtained from matrix $E_{z_j w_j}$ where the term $\x(t_{j+1},\cdot)$ is replaced by
$\widehat{\x}(t,\cdot)$ with $t\in [t_{j-1},\hspace*{1mm} t_{j}]$. Since $\widehat{\x}(t,\cdot)$ with $t\in [t_{j-1},\hspace*{1mm}
t_{j}]$ is known, matrix $E_{\hat{z}_j w_j}$ can be considered as an estimate of $E_{z_j w_j}$.

4. In the important case of an {\em additive} noise, $E_{z_j w_j}$  can be represented in the explicit form. Indeed, if
$$
\y(t,\cdot)= \x(t,\cdot)+ \xi(t,\cdot)
$$
where $\xi(t,\cdot)\in L^{2}(\Omega,{\mathbb R}^{m})$ is a random noise, then $\zz(t_j,t_{j+1},\cdot)=\y(t_{j+1},\cdot)
-\xi(t_{j+1},\cdot)-\widehat{\x}(t_j,\cdot)$ and matrix $E_{z_j w_j}$ can be represented  as follows:
\begin{eqnarray}\label{ezw11}
E_{z_j w_j} = E_{(y_{j+1}- \xi_{j+1})(y_{j+1}-y_{j})} - E_{{\small \widehat{x}_j}(y_{j+1}-y_{j})}
\end{eqnarray}
We note that  the RHS of (\ref{ezw11})  depends only on observed signals $\y(t_{j},\cdot)$, $\y(t_{j+1},\cdot)$,  estimated signal
$\widehat{\x}(t_{j},\cdot)$,
and noise $\xi(t_{j+1},\cdot)$, not on the  reference signal $\x(t_{j+1},\cdot)$. In particular, in (\ref{ezw11}), the term
$E_{\xi_{j+1}(y_{j+1}-y_{j})}$ can be estimated as
$\pm (E[\xi_{j+1}^2])^{1/2} (E[(y_{j+1}-y_{j})^2])^{1/2}$ where $ \displaystyle E[\xi_{j+1}^2]
=\int_{\Omega} [\xi(t_{j+1},\omega)]^2\ d\mu (\omega)$. It is motivated by the Holder's inequality for integrals.
 The second term in (\ref{ezw11}),
$E_{{\small \widehat{x}_j}(y_{j+1}-y_{j})}$, can be estimated from the samples of $\widehat{\x}(t_{j+1},\cdot)$ and
$\y(t_{j+1},\cdot)-\y(t_{j},\cdot)$.

We also note that the first term in the RHS of (\ref{ezw11}), $E_{(y_{j+1}- \xi_{j+1})(y_{j+1}-y_{j})}$, is
similar to the related covariance matrix in the Wiener filtering approach \cite{tor100}.

5. Other known ways to estimate $E_{\xi_{j+1}(y_{j+1}-y_{j})}$ can be found in \cite{tor100}, Section 5.3.
\medbreak

In general, an estimation of covariance matrices is a special research topic  which is not a subject
of this paper. The relevant references can be found, for example, in \cite{tor100,led1}.

\section{Simulations}\label{sim}


\subsection{General consideration}\label{gen1}

In  these simulations, in accordance with Section \ref{numer}, signal sets  $K_{_X}$ and $K_{_Y}$ (see Section \ref{not}) are
given by
$$
K_{_X}=\{\x(\tau_{1},\cdot), \x(\tau_{2},\cdot), \ldots, \x(\tau_{{N}},\cdot)\}\qa K_{_Y}=\{\y(\tau_{1},\cdot),
\y(\tau_{2},\cdot, \ldots, \y(\tau_{N},\cdot)\},
$$
 where, for $k=1,\ldots,N$, $\x(\tau_{k},\cdot)\in L^2(\Omega, \rt^m)$ and  $\y(\tau_{k},\cdot)
\in L^2(\Omega, \rt^n)$.    In many practical problems (arising, for example, in a DNA analysis
the number $N$ is quite large, for instance, $N=\mathcal O(10^4)$.

We set  $N=141$ and $m=n=116$. Thus, in these simulations, the interval $T=[a, \hspace*{1mm} b]$ (see Sections \ref{not} and \ref{numer}) is modelled as 141
points $\tau_{k}$ with $k=1,\ldots,141$ so that $[a, \hspace*{1mm} b] = [\tau_{1}, \tau_{2}, \ldots,  \tau_{141}]$.

The sequence of fixed time-points $\{t_k\}_1^p\subset T$  in (\ref{atjb}) is now such that
\begin{equation}\label{atjb1}
\tau_{1}= t_1 < \ldots < t_p = \tau_{141}.
\end{equation}
Below, in Examples 1-12, four particular choices of the specific interpolation signal pairs
$\{\x(t_{j},\cdot),$ $ \y(t_{j},\cdot)\}_1^p$ (introduced in Section \ref{inter})  are considered, for $p=5, 8, 15$ and $28$.
Points $t_1,\ldots, t_p$ are as follows.

For $p=5, 8, 15$, if $j=1,\ldots, p$, then $t_j =t^{(p)}_j = \tau_1 + (j-1) \Delta_p$,  respectively, where $\Delta_5 = 35,$
$\Delta_8 = 20$ and $\Delta_{15} = 10$.

For $p=28$, if $j=1,\ldots, p-1$, then $t_j =t^{(p)}_j = \tau_1 + (j-1) \Delta_{28}$, and if $j=p$, then $t_{28} =t^{(28)}_{28}
= t_{27} +  6 = 141$, where $\Delta_{28} = 5$.


Signals $\x(\tau_{k},\cdot)$ and  $\y(\tau_{k},\cdot)$ have been simulated as digital images represented by $116 \times 256$
matrices
\begin{eqnarray}\label{xkim}
X^{(k)}=[\x(\tau_{k},\omega_1), \ldots, \x(\tau_{k},\omega_{256})]\hspace{3mm}\mbox{and}\hspace{3mm}
Y^{(k)}=[\y(\tau_{k},\omega_1), \ldots, \y(\tau_{k},\omega_{256})],
\end{eqnarray}
 respectively, for $k=1,\ldots,141$, so that $X^{(k)}$ represents an image that should be estimated from an
observed image $Y^{(k)}$. A column of matrices $X^{(k)}$ and $Y^{(k)}$, $\x(\tau_{k},\omega_i)\in \rt^{116}$ and
$\y(\tau_{k},\omega_i)\in \rt^{116}$, for $i=1,\ldots,256$, represents a realization of
signals $\x(\tau_{k},\cdot)$ and $\y(\tau_{k},\cdot)$, respectively.

Note that $X^{(1)}, \ldots, X^{(141)}$ did not used in the piecewise linear  filter $F^{(p-1)}$ below  since they  are not
supposed to be known. They are represented here for illustration purposes only.
In particular, $X^{(1)}, \ldots, X^{(141)}$ are used to compare their estimates by different filters.

Observed noisy signals  $Y^{(1)}, \ldots, Y^{(141)}$ have been simulated in different forms presented by (\ref{sim1yk}),
(\ref{sim2yk}), (\ref{sim4yk}) and  (\ref{sim4yk2}) in the Examples 1-12 below. We note that the considered observed {\em signals are
grossly corrupted}.

To estimate the signals $X^{(1)}$, ..., $X^{(141)}$ from the  observed signals $Y^{(1)}$,  ..., $Y^{(141)}$, the proposed
piecewise linear filter $F^{(p-1)}$, the generic optimal linear (GOL) filters  $\cite{tor100}$ and  the averaging polynomial filter
$\cite{tor-m1}$ have been used.

The filters proposed in \cite{tor-m1,tor-m2,rus1,cou1} have not been applied here by the reasons discussed
in Section \ref{intr}. In particular, the filter  in \cite{cou1} cannot be applied  to signals represented by $Y^{(1)}$,
..., $Y^{(141)}$ in the form (\ref{sim1yk}), (\ref{sim2yk}), (\ref{sim4yk}) and  (\ref{sim4yk2}) below because
the associated inverse  matrices used in \cite{cou1} do not exist.

For signals under consideration (given by matrices $X^{(k)}$ and $Y^{(k)}$ with $k=1,\ldots, 141$), the filter  $F^{(p-1)}$, the
generic optimal linear (GOL) filters \cite{tor100} and the averaging polynomial filter $\cite{tor1,tor-m1}$ are represented as
follows.

(i) {\em Piecewise linear  filter $F^{(p-1)}$.}
 For $j=1,\ldots, p$, $\{X_j, Y_j\}$ designates an interpolation pair defined similarly to that in Section \ref{inter}. Each $X_j$ and  $Y_j$
 is associated with $t_j$ in (\ref{atjb1}) so that
 $$
X_j=[\x(t_j,\omega_1), \ldots, \x(t_j,\omega_{256})]\qa
Y_j=[\y(t_j,\omega_1), \ldots, \y(t_j,\omega_{256})].
$$
The estimate $\widehat{X}^{(k)}$ of $X^{(k)}$ by the filter $F^{(p-1)}$ is given by
\begin{equation}\label{fykk}
\widehat{X}^{(k)}=F^{(p-1)}[Y^{(k)}],
\end{equation}
where, by $(\ref{fyd0})$-$(\ref{b0j})$ in Section \ref{det1},
 \begin{eqnarray}\label{fykkj}
 && F^{(p-1)}[Y^{(k)}] = \sum_{j=1}^{p-1}\delta_j  F_j [Y^{(k)}],\quad \delta_j = \left\{ \begin{array}{cl}
                             1,  & \mbox{if $t_j\leq \tau_k \leq t_{j+1}$},\label{sim-f}\\
                             0, & \mbox{otherwise,}
                              \end{array} \right.\\
 &&F^{(p-1)}_j[Y^{(k)}] = \widehat{X}_j + B_j[Y^{(k)}-Y_j],\label{sim-fj}\\
       && \widehat{X}_j  = F_{j-1}[Y_j],\quad \mbox{$\widehat{X}_1$ is given},     \\
       && B_j = E_{{_Z}_j{_W}_{j}}(E_{{_W}_j{_W}_j})^{\dag},\label{sim-bj}
\end{eqnarray}
and where  $E_{{_Z}_j{_W}_{j}}$ and $E_{{_W}_j{_W}_j}$ are estimates of matrices $E_{z_{j}w_{j}}$ and $E_{w_{j}w_{j}}$ in
(\ref{b0j}),
 respectively. In particular, $E_{W_{j}W_{j}}$ can be represented  in the form
 \begin{equation}\label{ewwj}
 E_{{_W}_j{_W}_j} = W_jW^T_{j},\quad \mbox{where $W_j = Y_{j+1}-Y_{j}$.}
\end{equation}

 Further, matrix $E_{{_Z}_j{_W}_{j}}$ depends on $Z_j =X_{j+1} -  \widehat{X}_j$ where $X_{j+1}$ is unknown.
 Therefore a determination of $E_{{_Z}_j{_W}_{j}}$ is reduced, in fact, to finding an estimate of $X_{j+1}$.
 Since it is customary to find
 $E_{{_Z}_j{_W}_{j}}$ in terms of signal samples \cite{tor100}, $E_{{_Z}_j{_W}_{j}}$ has been presented as
  \begin{equation}\label{ezjzj}
 E_{{_Z}_j{_W}_{j}}=\widetilde{Z}_jW^T_j,\quad \mbox{where  $\widetilde{Z}_j=\widetilde{X}_{j+1}- \widehat{X}_j$}
 \end{equation}
  and
 $\widetilde{X}_{j+1}$ has been constructed from a sample of $X_{j+1}$ as follows. The sample of $X_{j+1}$  is
 a $116\times 128$ matrix  presented by odd columns of
$X_{j+1}$. Then an estimate of $X_{j+1}$ is chosen as a $116\times 256$ matrix $\widetilde{X}_{j+1}$ where each odd column is
a related odd column of  $X_{j+1}$, and each even column is an average of two
 adjacent columns. The last column in   $\widetilde{X}_{j+1}$ is the same as its preceding column.

This way of estimating $E_{{_Z}_j{_W}_{j}}$ was chosen   for illustration purposes only. Other related methods have been considered
in Section \ref{some}.


The errors associated with the filter $F^{(p-1)}$ are given by
  \begin{equation}\label{varf1}
 \varepsilon_{k, F}^{(p-1)} =  \left \|X^{(k)} - {F}^{(p-1)}[Y^{(k)}]\right \|^2_F,\quad \mbox{for $k=1,\ldots,141$}.
\end{equation}

(ii) {\em Generic optimal linear (GOL) filters} $\cite{tor100}$. To each signal $Y^{(k)}$, an individual
GOL filter $W_k$ has also been applied, so that $W_k$ estimates $X^{(k)}$ from $Y^{(k)}$ in the form
$$
W_k Y^{(k)} = E_{X^{(k)}Y^{(k)}}E_{Y^{(k)}Y^{(k)}}^\dag Y^{(k)},
$$
for each $k=1,\ldots,141$. Thus, the GOL filter $W_k$ requires an estimate of 141 matrices  $E_{X^{(k)}Y^{(k)}}$, for each
$k=1,\ldots, 141$.

Similarly to matrix $E_{Z_jW_j}$ in the filter $F^{(p-1)}$ above,
the matrix $E_{X^{(k)}Y^{(k)}}$ has been estimated from samples of each $X^{(k)}$,  $\widetilde{X}^{(k)}$, for each
$k=1,\ldots, 141$.

{\em One of the advantages
of the proposed
filter  $F^{(p-1)}$} is that $F^{(p-1)}$ requires a smaller number, $p$, of samples of $X_j$, $\widetilde{X}_j$, to be known
(where $j=1,\ldots, p$).

The errors associated with filters $W_k$ are given by
 \begin{equation}\label{varw2}
\displaystyle \epsilon_{k,w} = \|X^{(k)} - {W_k}Y^{(k)}\|^2_F.
\end{equation}

(iii) {\em Averaging polynomial filters} $\cite{tor1,tor-m1}$.
By the methodology in \cite{tor1}, the averaging polynomial filter $W$ is based on the use of the estimates of
the covariance matrices, $E_{XY}$
 and $E_{YY}$, in the form
$$
E_{XY} =\frac{1}{141}\sum_{k=1}^{141} \widetilde{X}^{(k)}(Y^{(k)})^T \qa E_{YY} =\frac{1}{141}\sum_{k=1}^{141} Y^{(k)}(Y^{(k)})^T.
$$
Then, for each, $k=1,\ldots,141$, the estimate of $X^{(k)}$ is given by
$$
W Y^{(k)} = E_{XY}E_{Y Y}^\dag Y^{(k)}.
$$
The errors associated with the filter $W$ are given by
 \begin{equation}\label{varw1}
\displaystyle \varepsilon_{k_W} = \|X^{(k)} - {W}Y^{(k)}\|^2_F,\quad \mbox{for $k=1,\ldots,141$}.
\end{equation}

\subsection{Simulations with signals modeled from images `plant': application  of piecewise interpolation filter
and GOL filters}\label{sim0}

Here, results of simulations for reference signals represented by matrices $X^{(1)},$
$\ldots,$  $X^{(141)}$   (see ({\ref{xkim}) above) formed from images
`plant'\footnote{The database is available in http://sipi.usc.edu/services/database.html.} are considered.
 Typical selected images $X^{(k)}$  are shown in Fig. \ref{fig1}.

Observed noisy images $Y^{(1)}, \ldots, Y^{(141)}$ have been simulated in the form
\begin{equation}\label{sim1yk}
Y^{(k)} = X^{(k)}\bullet \mbox{\tt randn}_{(k)}\bullet \mbox{\tt rand}_{(k)},
\end{equation}
for each $k=1,\ldots,141.$
Here, $\bullet$ means the Hadamard product, and $\mbox{\tt randn}_{(k)}$ and $\mbox{\tt rand}_{(k)}$ are $116\times 256$
matrices with random entries. The entries of $\mbox{\tt randn}_{(k)}$ are  normally distributed with mean zero, variance
one and standard deviation one.  The entries of $\mbox{\tt rand}_{(k)}$  are  uniformly distributed in the interval
$(0, 1)$. A typical example of such images is given in Fig. \ref{fig2} (a).

To demonstrate the effectiveness of the proposed filter $F^{(p-1)}$,  sub-filters $F^{(p-1)}_j$ and  associated
interpolation signal pairs $\{X_j, Y_j\}_{j=1}^p$  have been chosen in four different ways as follows.

{\em Example 1}. First, for $p=5$, the interpolation signal pairs are
\begin{eqnarray}\label{pair1}
&&\hspace*{-10mm}\{X_1, Y_1\}=\{X^{(1)}, Y^{(1)}\},\quad \{X_2, Y_2\}=\{X^{(35)}, Y^{(35)}\},\quad
 \{X_3, Y_3\}=\{X^{(70)}, Y^{(70)}\},\label{sim1x1} \\
&&\hspace*{10mm} \{X_4, Y_4\}=\{X^{(105)}, Y^{(105)}\},\quad  \{X_5, Y_5\}=\{X^{(141)}, Y^{(141)}\}.\label{sim1x4}
\end{eqnarray}
The error values $\{\varepsilon_{k, F}^{(4)}\}_1^{141}$  associated with filter $F^{(4)}$ are evaluated  by (\ref{varf1}).
The graph of $\{\varepsilon_{k, F}^{(4)}\}_1^{141}$  is presented in  Fig. \ref{fig3} (a).

{\em Example 2}. For $p=8$, the interpolation signal pairs are
\begin{eqnarray}\label{pair2}
&&\hspace*{-10mm}\{X_1, Y_1\}=\{X^{(1)}, Y^{(1)}\}, \quad \{X_j, Y_j\}
=\{X^{(20(j-1))}, Y^{(20(j-1))}\},\quad \mbox{for $j=2,\ldots,7$};\label{sim1x21} \\
&&\hspace*{30mm}\mbox{and}\quad \{X_8, Y_8\}=\{X^{(141)},Y^{(141)}\}.\label{sim1x8}
\end{eqnarray}
The error magnitudes $\{\varepsilon_{k, F}^{(7)}\}_1^{141}$ associated  with the piecewise interpolation filter $F^{(7)}$
constructed by (\ref{sim-f})-(\ref{ezjzj}) with the interpolation signal pairs given by (\ref{sim1x21})-(\ref{sim1x8})
 are diagrammatically shown  in Fig. \ref{fig3} (b).

It follows from Fig.  \ref{fig3} (b) that the errors associated with filter $F^{(7)}$ is less than those of filter $F^{(4)}$.
This is a confirmation of Theorem \ref{th3}.

{\em Example 3}. Further, for$p=15$, the interpolation pairs are
\begin{eqnarray}\label{pair3}
&&\hspace*{-10mm} \{X_1, Y_1\}=\{X^{(1)}, Y^{(1)}\}, \quad \{X_j, Y_j\}=
\{X^{(10(j-1))},Y^{(10(j-1))}\} \quad \mbox{for $j=2,\ldots,14$};\label{sim1x31} \\
&&\hspace*{30mm}\mbox{and}\quad \{X_{15}, Y_{15}\}=\{X^{(141)},Y^{(141)}\}. \label{sim1x15}
\end{eqnarray}
In Fig.  \ref{fig3} (c), the errors  $\{\varepsilon_{k, F}^{(15)}\}_1^{141}$ associated  with the piecewise interpolation filter
$F^{(15)}$ are presented.  The Fig.  \ref{fig3} (c) demonstrates a further confirmation of Theorem \ref{th3}: the errors
associated with the piecewise interpolation filter diminishes as $p$ increases.

{\em Example 4}. Finally, the number  of interpolation signal pairs $\{X_j, Y_j\}_{j=1}^p$ is $p=29$ so that
\begin{eqnarray}\label{pair4}
&&\hspace*{-10mm} \{X_1, Y_1\}=\{X^{(1)}, Y^{(1)}\}, \quad \{X_j, Y_j\}
=\{X^{(5(j-1))}, \quad Y^{(5(j-1))}\} \quad \mbox{for $j=2,\ldots,28$};\label{sim1x41} \\
&&\hspace*{30mm}\mbox{and}\quad  \{X_{29}, Y_{29}\}=\{X^{(141)},Y^{(141)}\}.\label{sim1x29}
\end{eqnarray}

In this case, when $p$ is grater than in the previous Examples 1-3, the errors  $\{\varepsilon_{k, F}^{(29)}\}_1^{141}$
associated  with the piecewise interpolation filter $F^{(29)}$ are smaller than those associated with  filters $F^{(4)}$,
$F^{(8)}$ and $F^{(15)}$ - see Fig. \ref{fig3} (d).
\\

The diagrams   of errors associated with the GOL filters \cite{tor100} are also presented in Fig. \ref{fig3}.
It follows from Fig. \ref{fig3}  that proposed filters $F^{(4)}$, $F^{(8)}$, $F^{(15)}$ and $F^{(29)}$ provide the  better
accuracy then that of the GOL filters.

At the same time, the filter  $F^{(p-1)}$ is easer to implement since it requires less initial information  compared to  GOL filters,
as it has been discussed in Sections \ref{mot1} and \ref{comp}.

Results of the application of the averaging polynomial filter \cite{tor1,tor-m1} are discussed in Section \ref{sim3} below.

\subsection{Simulations with signals modelled from images `boat': application  of piecewise interpolation filter
and GOL filters}\label{boat2}

In this section, results of the simulations for a different type of signals than those
considered in Section \ref{sim0} above are presented. Here, the reference signals $X^{(1)},$ $\ldots,$  $X^{(141)}$ are formed from images
`boat'\footnote{The database is available in http://sipi.usc.edu/services/database.html.}.

Observed noisy signals $Y^{(1)}, \ldots, Y^{(141)}$ have been simulated in the form
\begin{equation}\label{sim2yk}
Y^{(k)} = X^{(k)}\bullet \mbox{\tt randn}_{(k)},
\end{equation}
for each $k=1,\ldots,141.$ The noise term is different from that in (\ref{sim1yk}).

 Typical selected images $X^{(k)}$ and $Y^{(k)}$ are shown in Figs. \ref{fig4} and \ref{fig5}, respectively.

As in Section \ref{sim0}, the piecewise interpolation filter $F^{(p-1)}$
is constructed by (\ref{sim-f})-(\ref{ezjzj}). In Examples 5-8 below, the  number $p-1$ of sub-filters $F^{(p-1)}_j$ and
associated interpolation signal pairs $\{X_j, Y_j\}_{j=1}^p$  have been chosen in four different ways.

{\em Example 5.} First, similar to Example 1, the number of interpolation signal pairs $\{X_j,Y_j\}_{j=1}^p$  has been
chosen
as $p=5$, and $X_j$ and $Y_j$ have been presented as in (\ref{pair1})-(\ref{sim1x4}).

The error values $\{\varepsilon_{k, F}^{(4)}\}_1^{141}$ associated with  the piecewise interpolation filter $F^{(4)}$
applied to these data are  presented  in  Fig. \ref{fig6} (a).

{\em Example 6.} For the grater number of interpolation signal pairs than that in Example 5, $p=8$, and for $X_j$ and $Y_j$
($j=1,\ldots, 8$) chosen as in  (\ref{pair2})-(\ref{sim1x8}),  the error magnitudes $\{\varepsilon_{k, F}^{(7)}\}_1^{141}$
associated  with the piecewise
interpolation filter $F^{(7)}$  are diagrammatically shown  in Fig. \ref{fig6} (b).
A comparison between Figs. \ref{fig6} (a) and (b) demonstrates that the increase in $p$ implies the decrease in the errors
associated with the filter
$F^{(p-1)}$.

{\em Example 7.} For $p=15$, and for $X_j$ and $Y_j$ ($j=1,\ldots, 15$) chosen as in (\ref{pair3})-(\ref{sim1x15}), the
errors
$\{\varepsilon_{k, F}^{(14)}\}_1^{141}$ associated  with the piecewise
interpolation filter $F^{(14)}$ are further less than those for filters $F^{(4)}$ and $F^{(7)}$. See Fig. \ref{fig6} (c) in this
regard.

{\em Example 8.} The further increase in $p$ to  $p=29$, confirms this tendency. The piecewise interpolation filter
$F^{(28)}$ with $X_j$ and $Y_j$ ($j=1,\ldots, 29$) chosen similar to (\ref{pair4})-(\ref{sim1x29}) produces the associated
errors
$\{\varepsilon_{k, F}^{(28)}\}_1^{141}$ represented in Fig.  \ref{fig6} (d). They are, clearly,  less than the errors
associated with  filters $F^{(4)}$,  $F^{(7)}$ and $F^{(15)}$.

\medbreak
The errors associated with the GOL filters are also presented in Figs. \ref{fig6} (a)-(d). The figures
clearly demonstrate the advantage of the  piecewise interpolation filter $F^{(p-1)}$.

Results of the application of the averaging polynomial filter \cite{tor-m1} are discussed in Section \ref{sim3} below.

\subsection{Results of simulations for averaging polynomial filter \cite{tor1,tor-m1}}\label{sim3}

To further illustrate the effectiveness of the proposed piecewise interpolation filter, in this Section, results of simulations for the averaging
polynomial filter \cite{tor1,tor-m1} are presented.
The filter has been applied to two different types of data considered in Sections \ref{sim0} and \ref{boat2}.

{\em Example 9.} The filter \cite{tor1,tor-m1} applied to signals considered in Section \ref{sim0} gives the associated errors
$\{\epsilon_{k_W}\}_{k=1}^{141}$  (see (\ref{varw1})) represented in Fig. \ref{fig7} (a). For a comparison, the errors
associated with the  piecewise interpolation filter $F^{(28)}$
and the GOL filters \cite{tor100} are also given in Fig.  \ref{fig7} (a).

A typical example of the estimated signal by the
averaging polynomial filter \cite{tor1,tor-m1} is presented in Fig. \ref{fig2} (d) above.

{\em Example 10.} The averaging polynomial filter \cite{tor-m1} applied to signals considered in Section \ref{boat2}
produces the associated errors $\{\epsilon_{k_W}\}_{k=1}^{141}$  shown in Fig. \ref{fig7} (b). The errors  associated
with the  piecewise interpolation filter $F^{(28)}$ are much smaller and they are not discerned in Fig. \ref{fig7} (b).
\medbreak

Together with Figs. \ref{fig2}, \ref{fig3}, \ref{fig5} and \ref{fig6}, Figs. \ref{fig7} (a) and (b) illustrate the advantage
of the piecewise interpolation filter.

\begin{figure}[]
\centering
 \hspace*{-7mm}\begin{tabular}{c@{\hspace*{0mm}}c}
 \includegraphics[width=8cm, height=3.7cm]{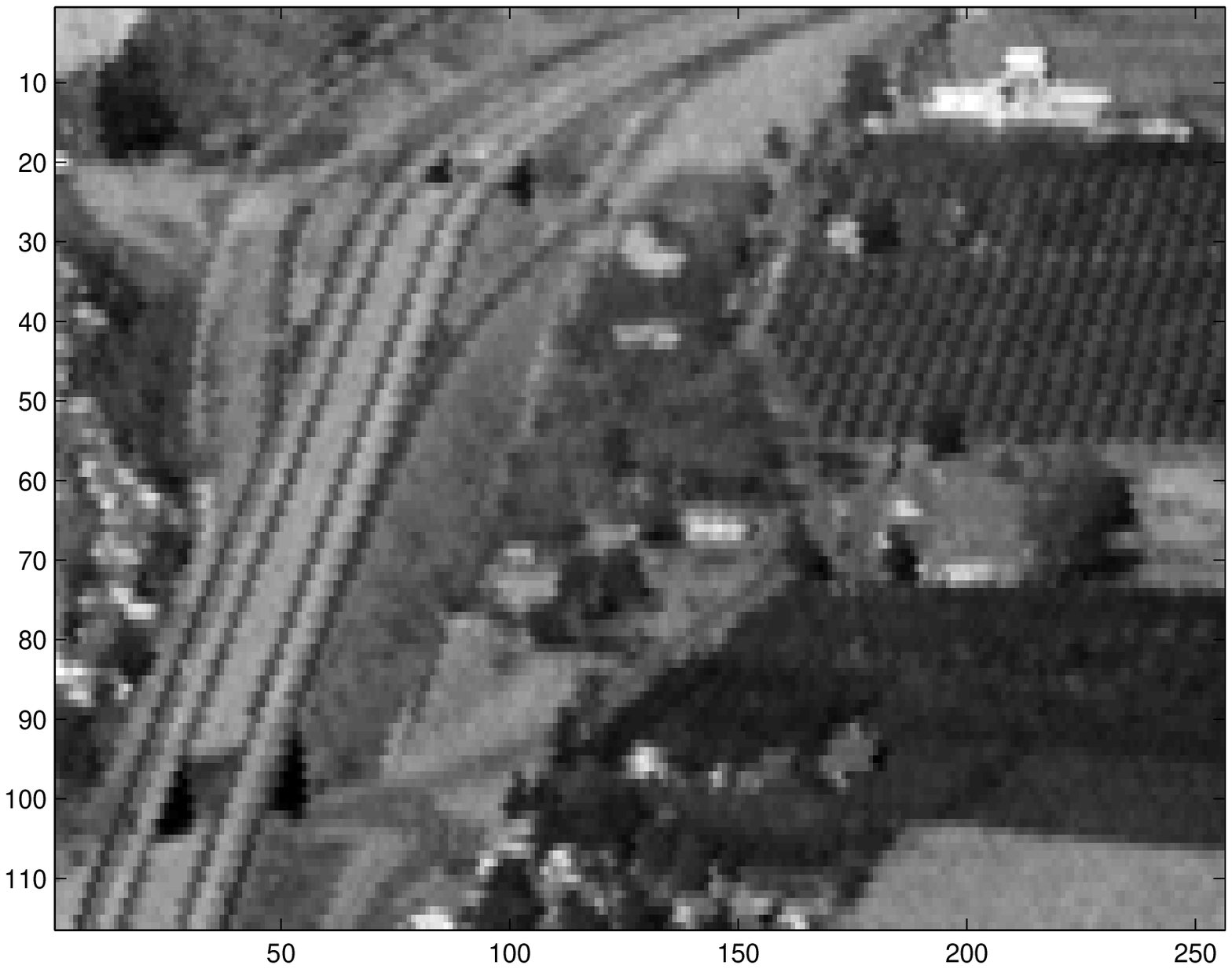} &  \includegraphics[width=8cm, height=3.7cm]{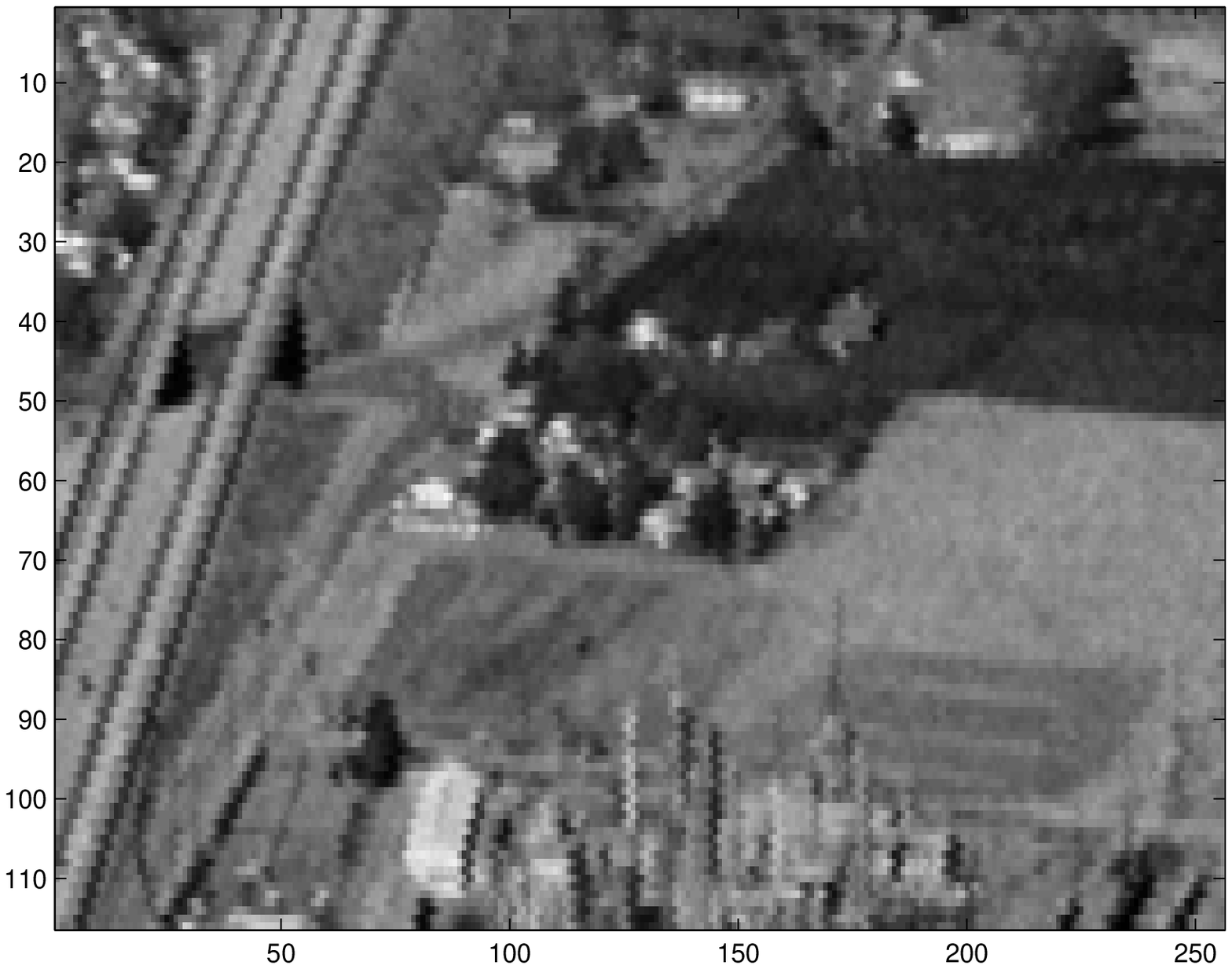}\\
{\small (a)  Signal $X^{(1)}.$ } & {\small (b) Signal $X^{(55)}.$} \\
  \includegraphics[width=8cm, height=3.7cm]{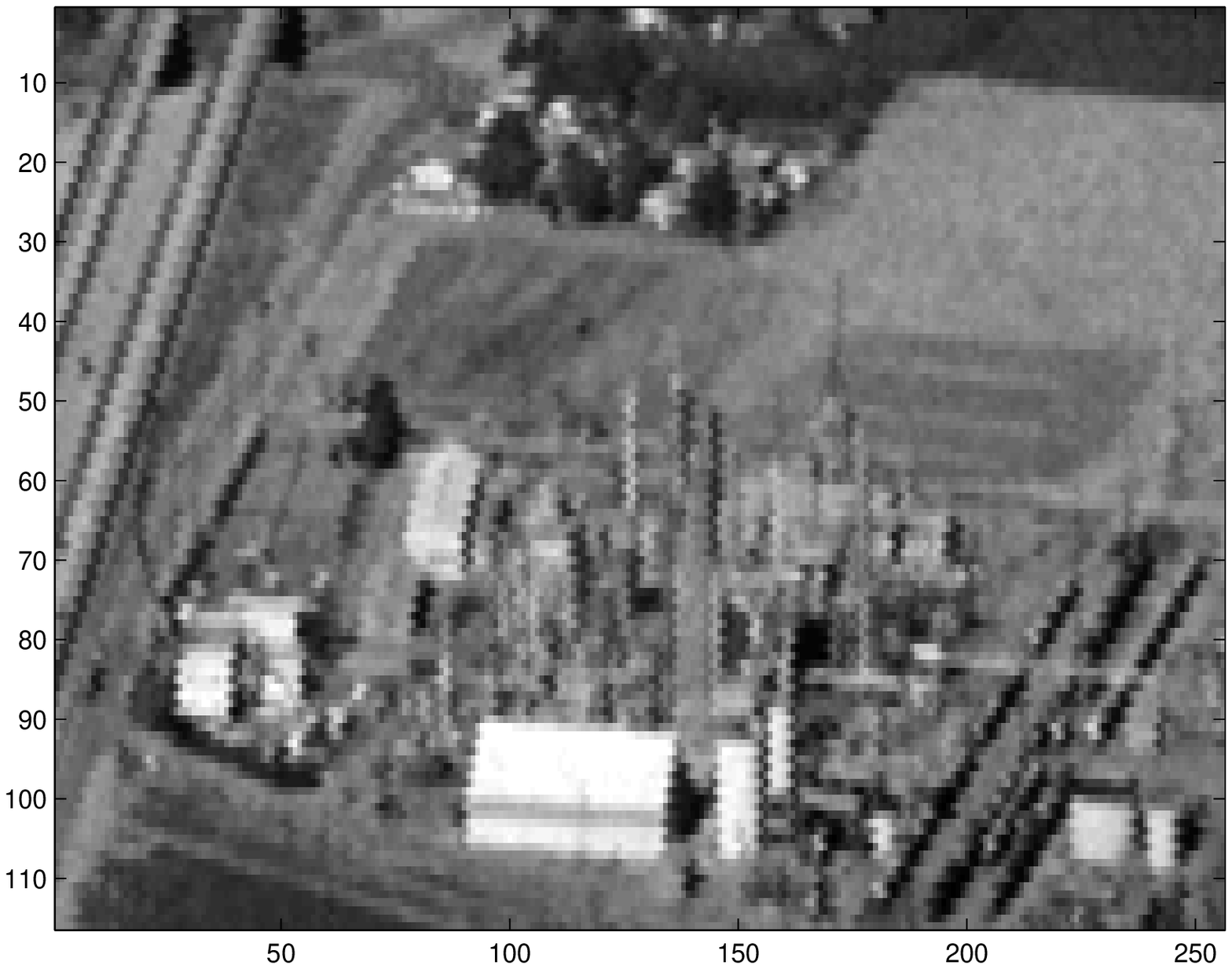}  &  \includegraphics[width=8cm, height=3.7cm]{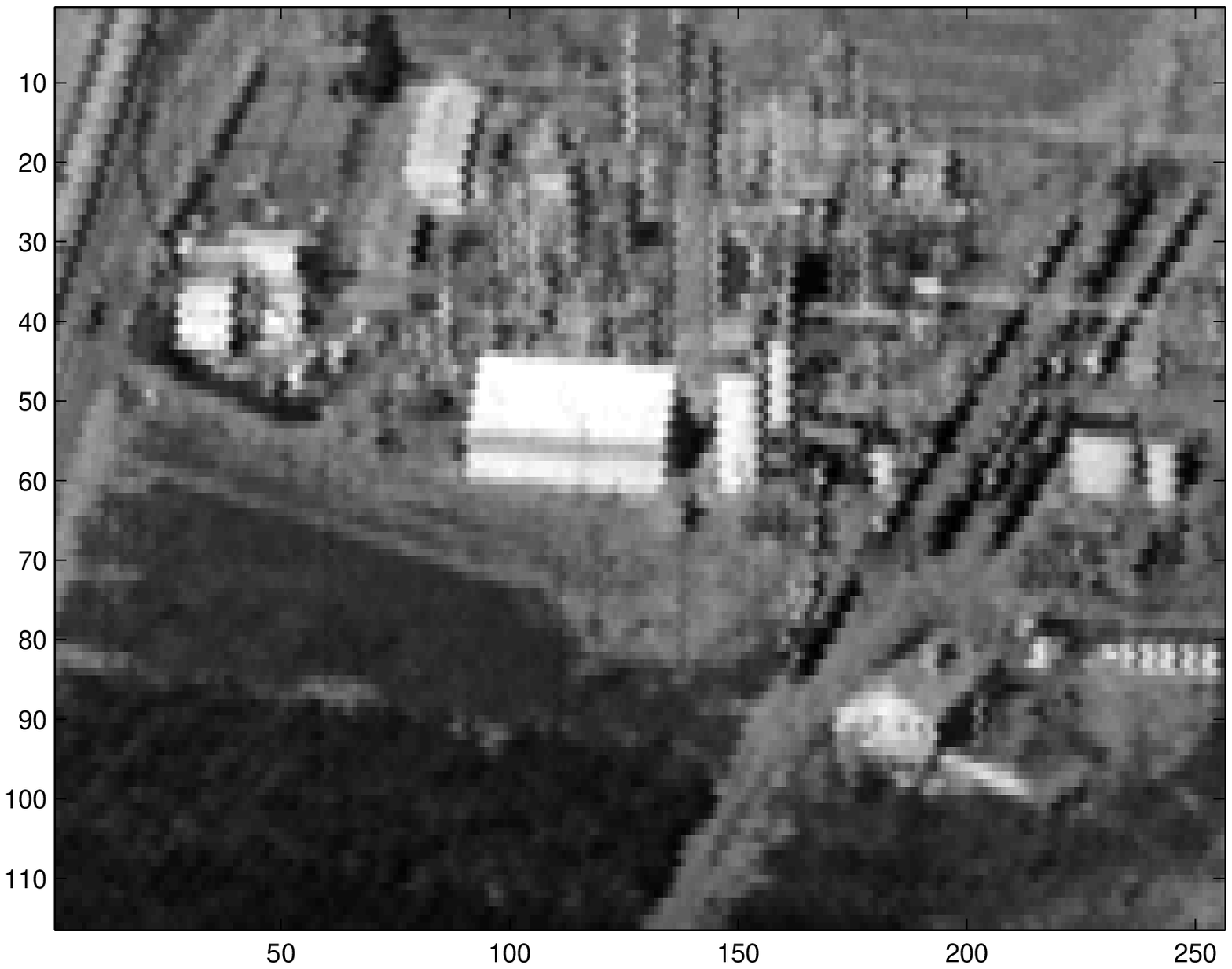}\\
 {\small   (c)  Signal $X^{(95)}.$}  & {\small (d) Signal $X^{(141)}.$ }
\end{tabular}
 \vspace*{0mm}\caption{Examples of selected signals  to be estimated from observed data.}
 \label{fig1}
 \end{figure}

\begin{figure}[]
\centering
 \hspace*{-7mm}\begin{tabular}{c@{\hspace*{0mm}}c}
  \includegraphics[width=8cm, height=3.7cm]{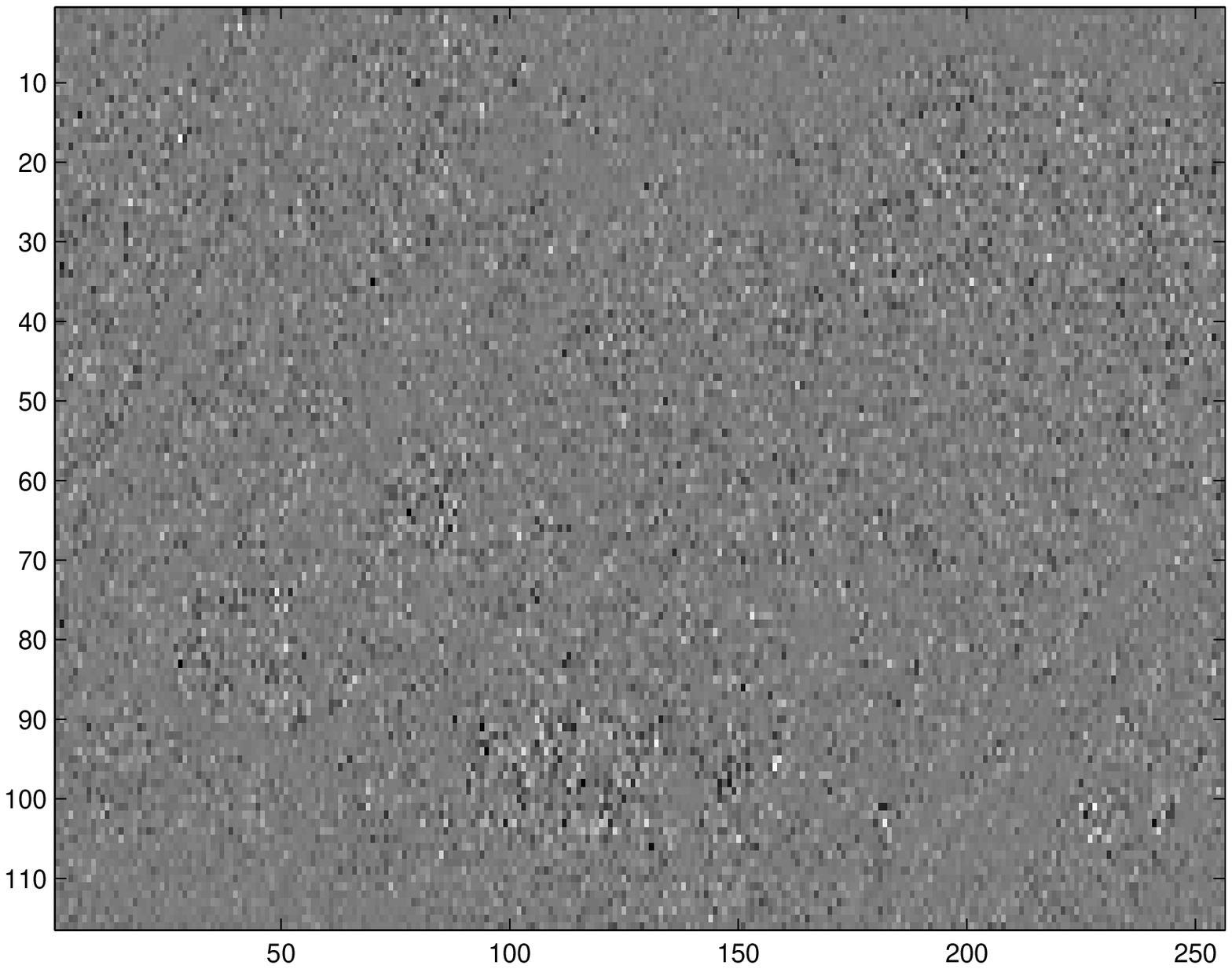} &  \includegraphics[width=8cm, height=3.7cm]{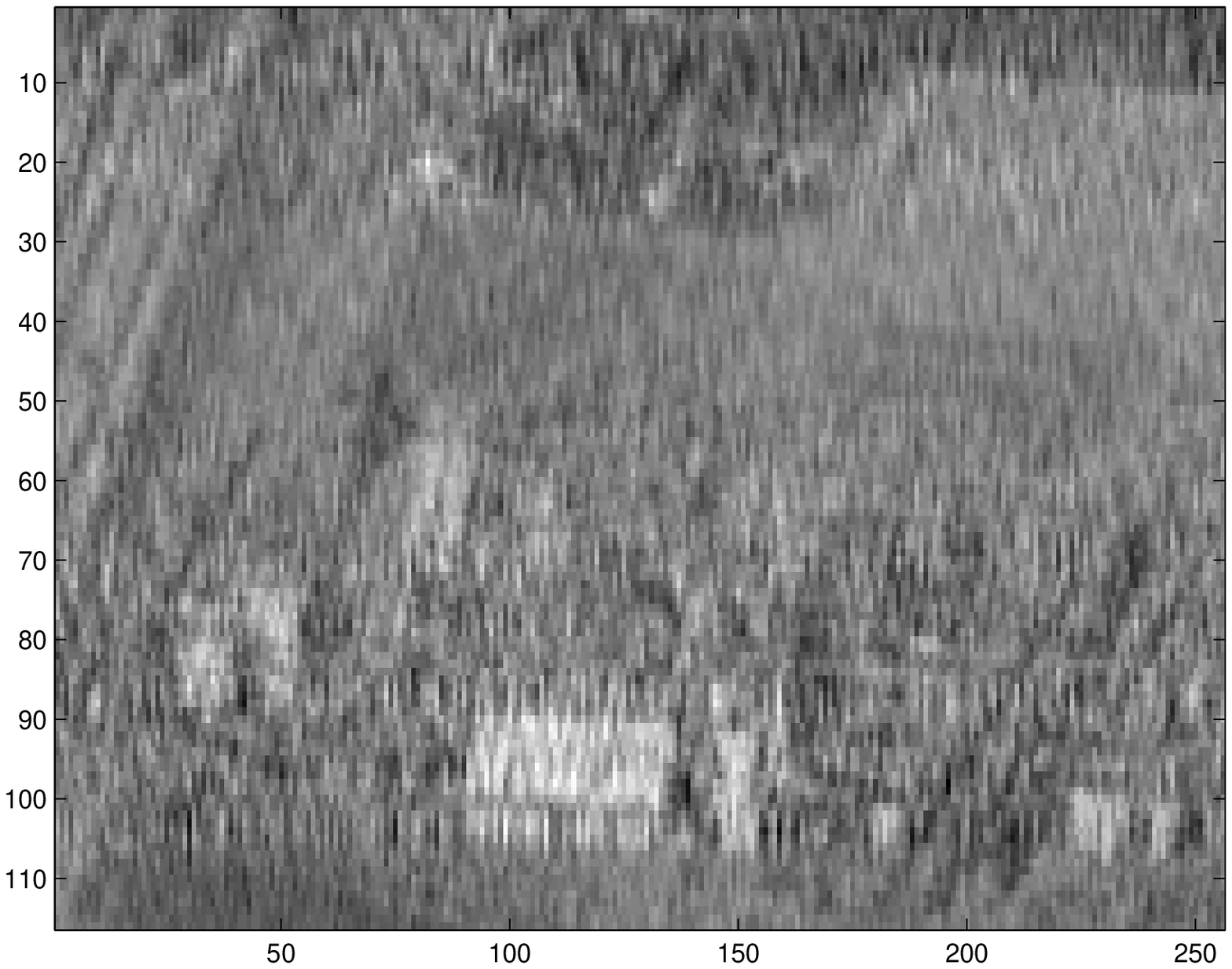}\\
   {\small   (a) Observed signal  $Y^{(95)}.$} &  {\small   (b) Estimate of $X^{(95)}$ by piecewise filter $F^{(28)}$.} \\
     \includegraphics[width=8cm, height=3.7cm]{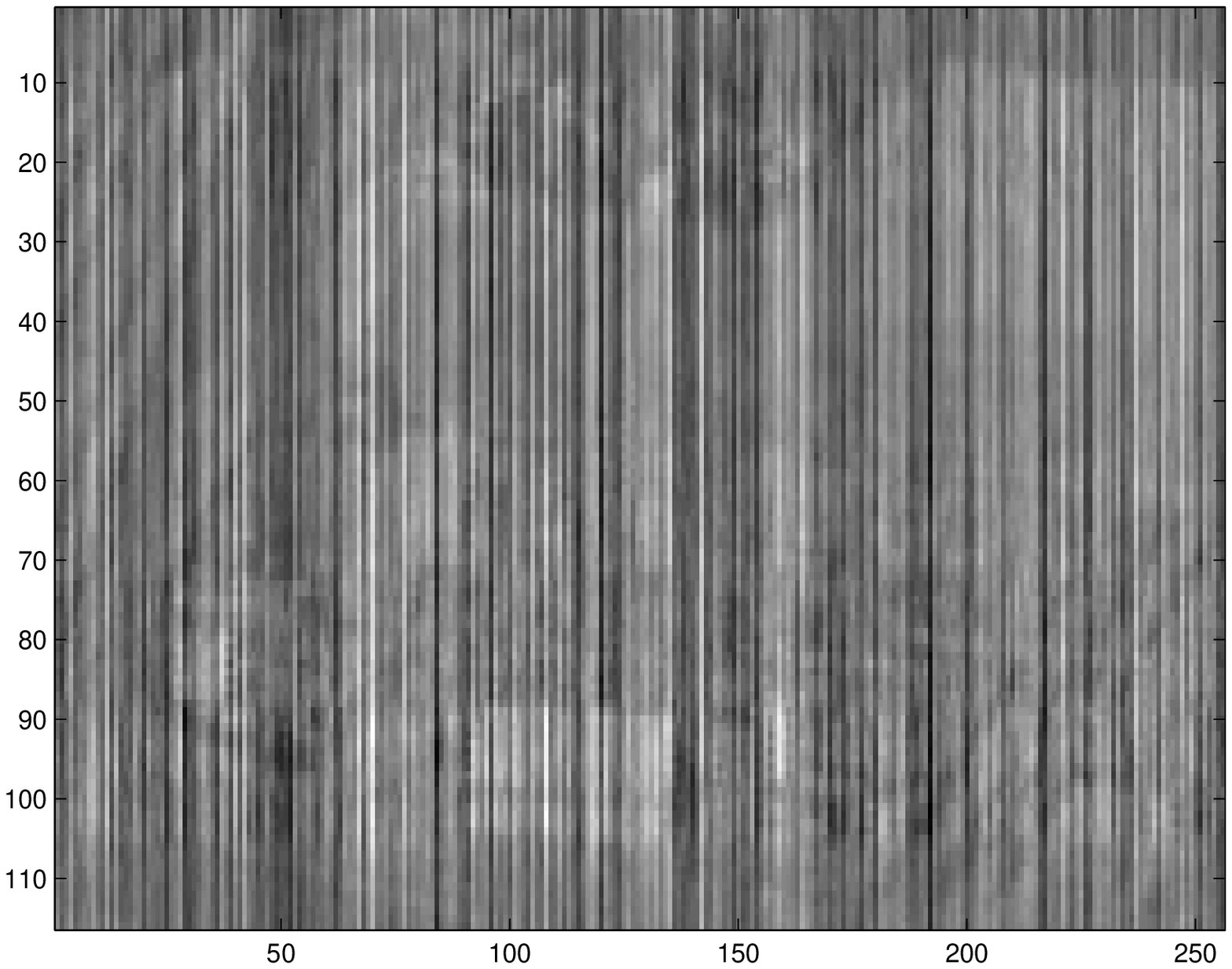} &  \includegraphics[width=8cm, height=3.7cm]{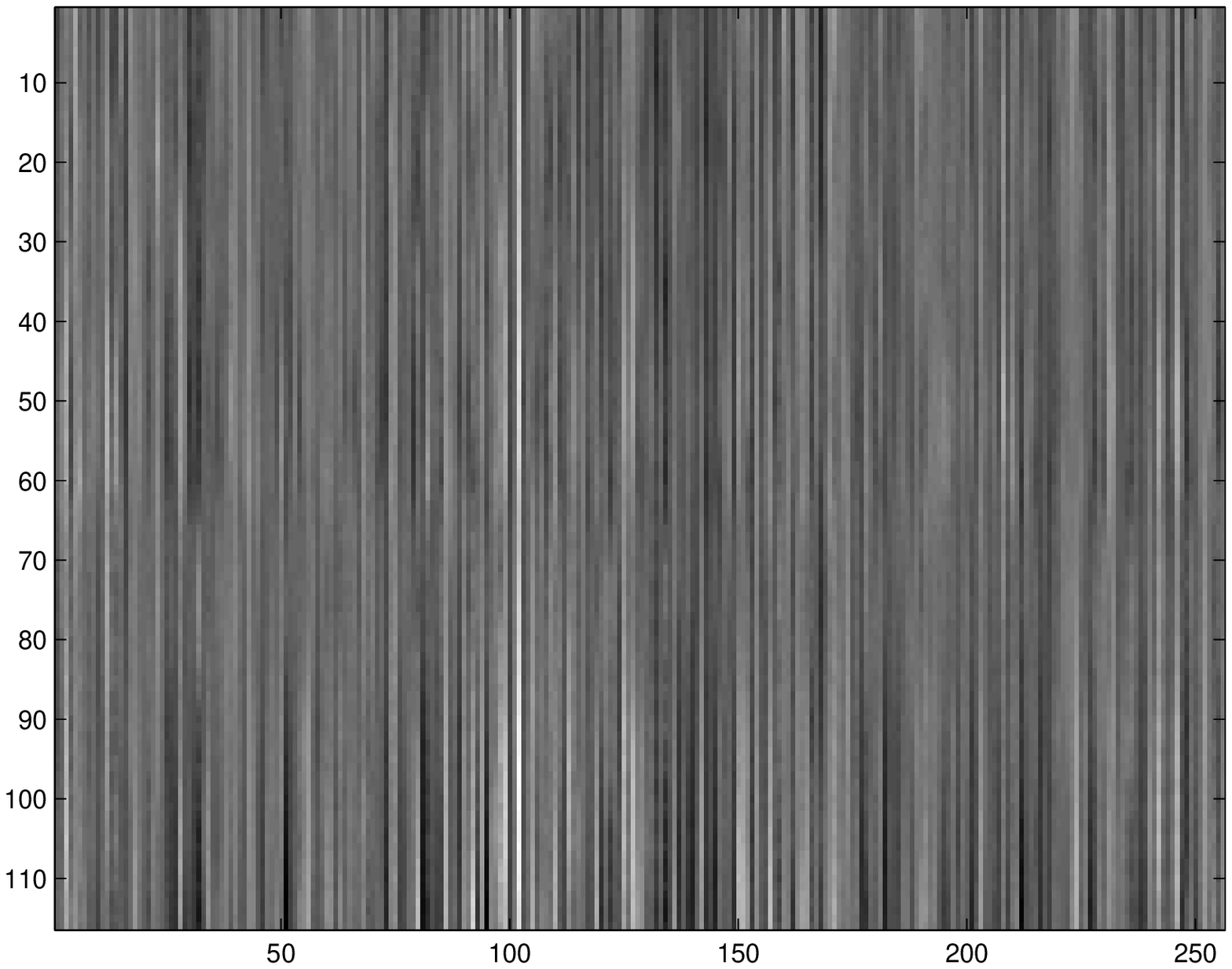} \\
   {\small   (c)  Estimate of $X^{(95)}$  by generic optimal linear  filter.} &   {\small   (d) Estimate of $X^{(95)}$ by averaging polynomial filter.}
   
\end{tabular}
 \vspace*{-5mm}\caption{Examples of the observed signal and the  estimates obtained by different filters.}
 \label{fig2}
 \end{figure}

\hspace*{-10mm}\begin{figure}[]
\centering
 \vspace*{-5mm}\begin{tabular}{c@{\hspace*{5mm}}c}
   \includegraphics[width=7.5cm, height=3.7cm]{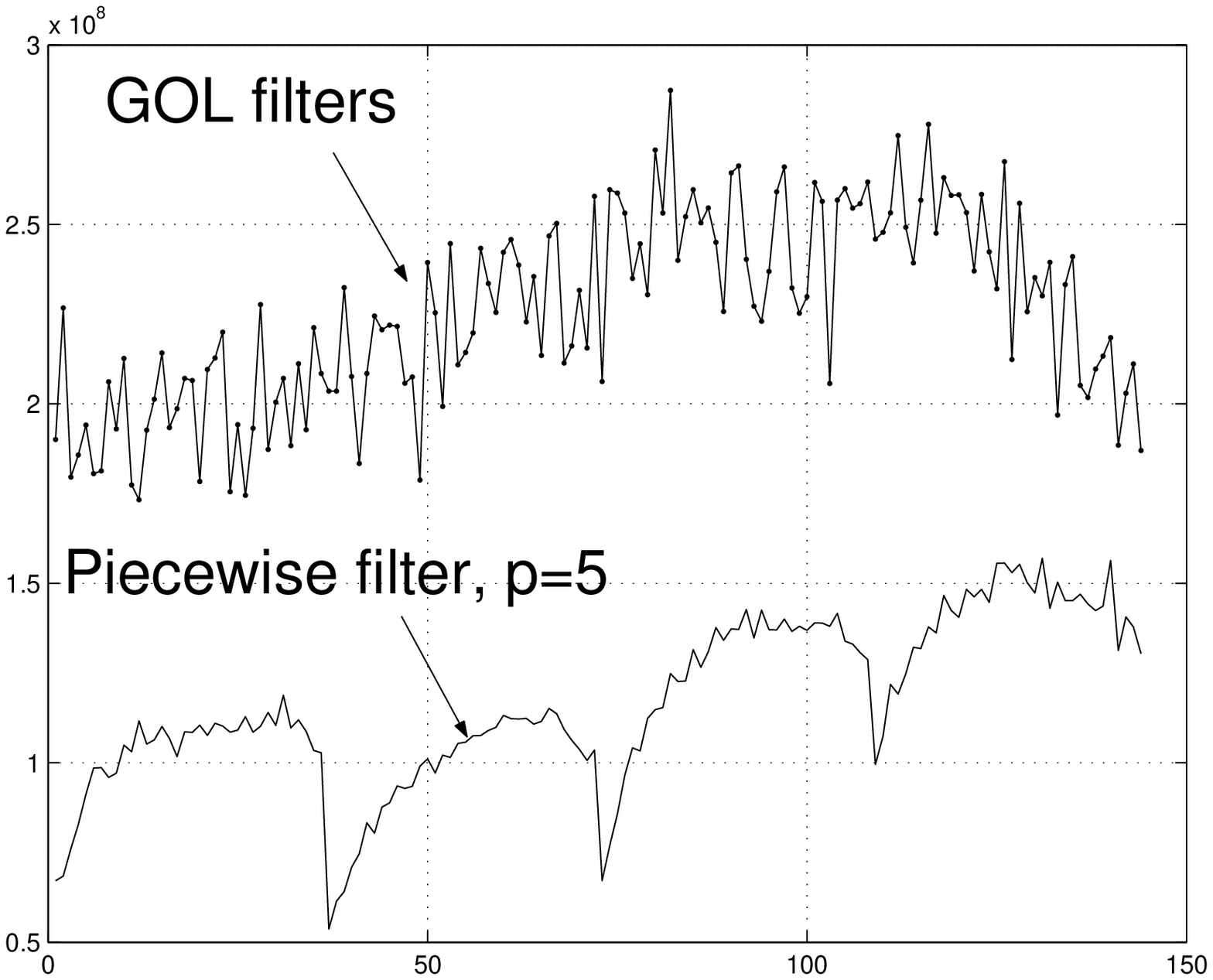} &  \includegraphics[width=7.5cm, height=3.7cm]{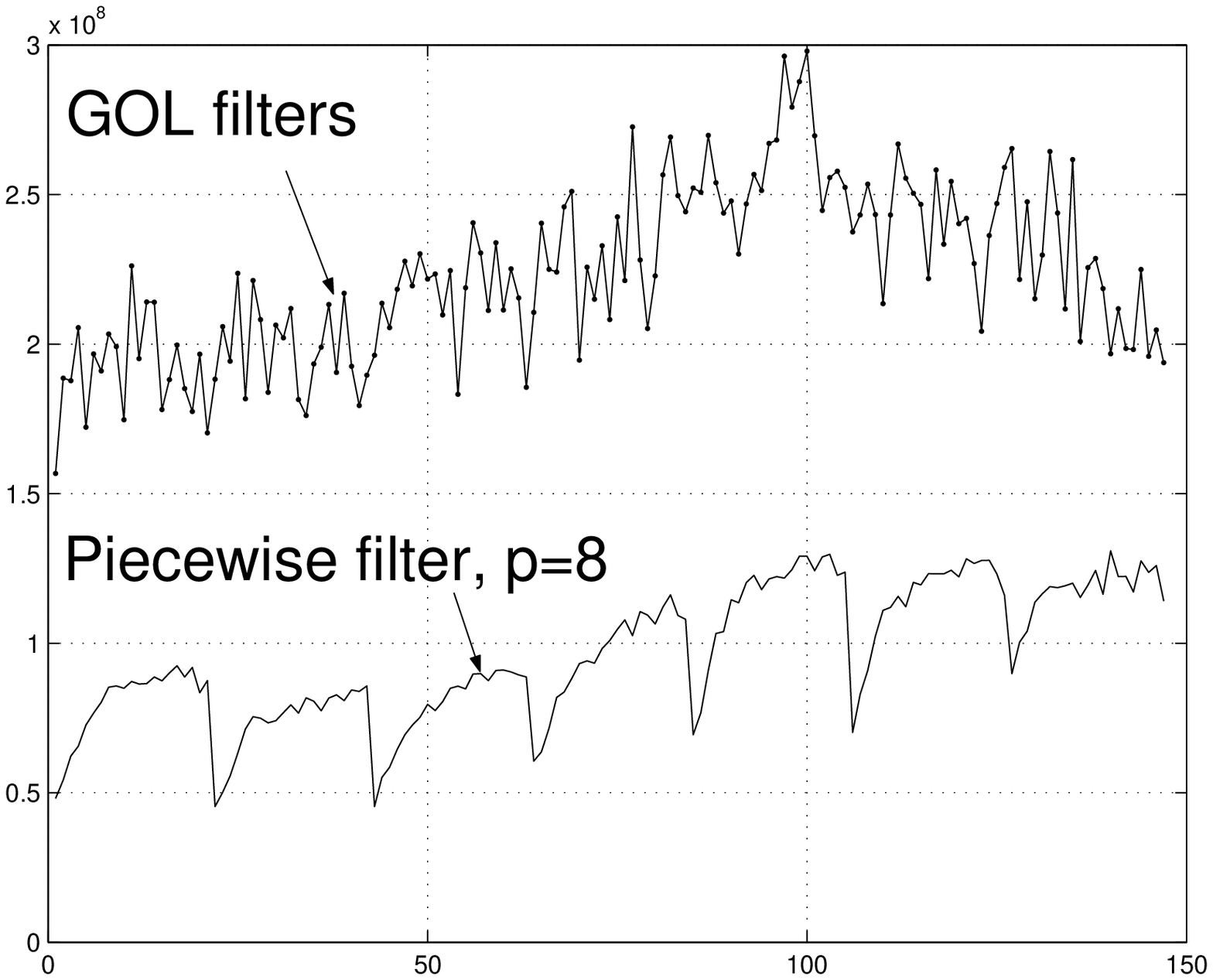}\\ 
    \includegraphics[width=7.5cm, height=3.7cm]{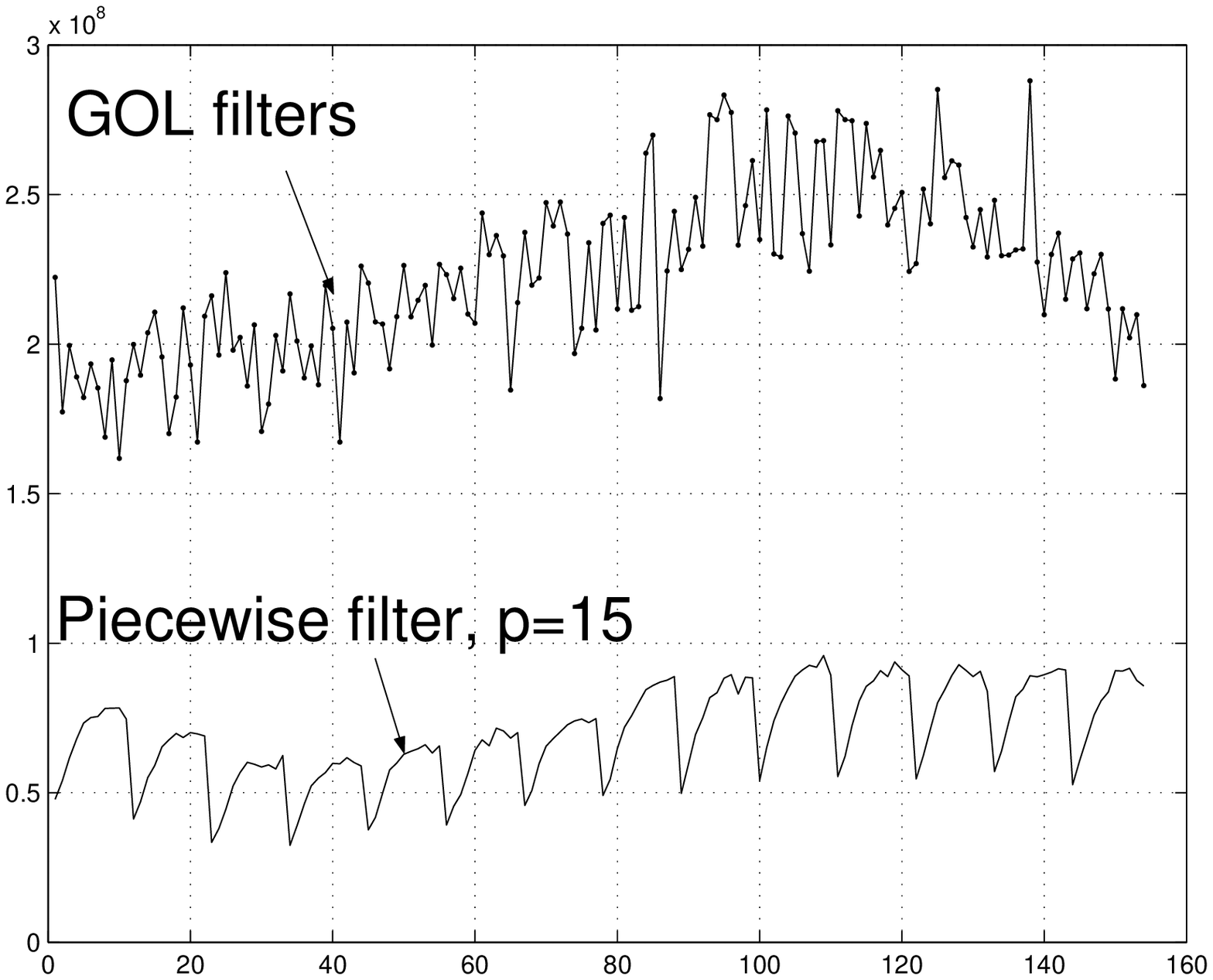} &  \includegraphics[width=7.5cm, height=3.7cm]{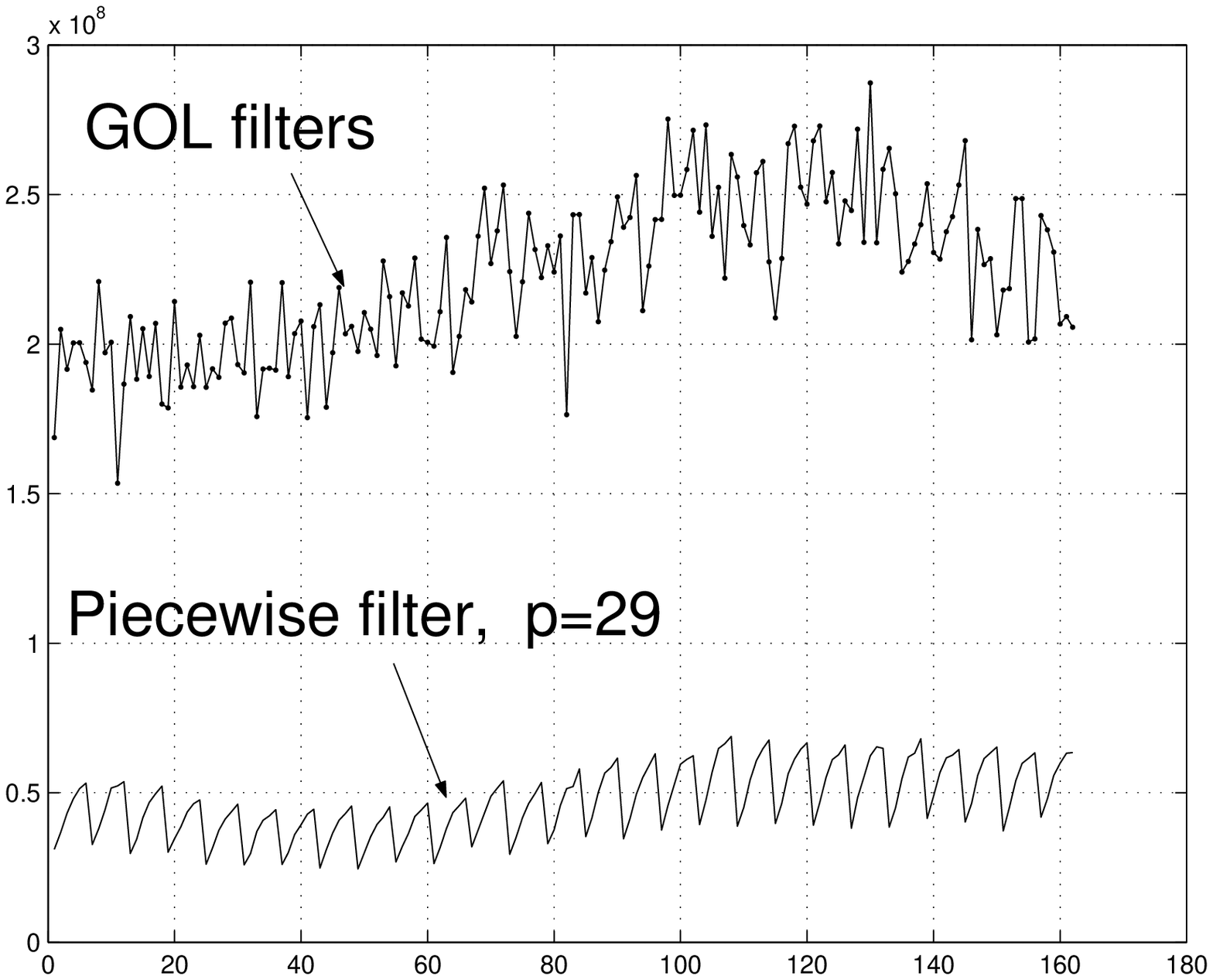}
\end{tabular}
 \vspace*{2mm}\caption{Illustration of the errors associated with the piecewise interpolation filters $F^{(p-1)}$
  and the generic optimal linear (GOL) filters \cite{tor100} applied to signals described in Examples 1--4.}
 \label{fig3}
 \end{figure}

\begin{figure}[]
\centering
 \hspace*{-7mm}\begin{tabular}{c@{\hspace*{0mm}}c}
   \includegraphics[width=8cm, height=3.7cm]{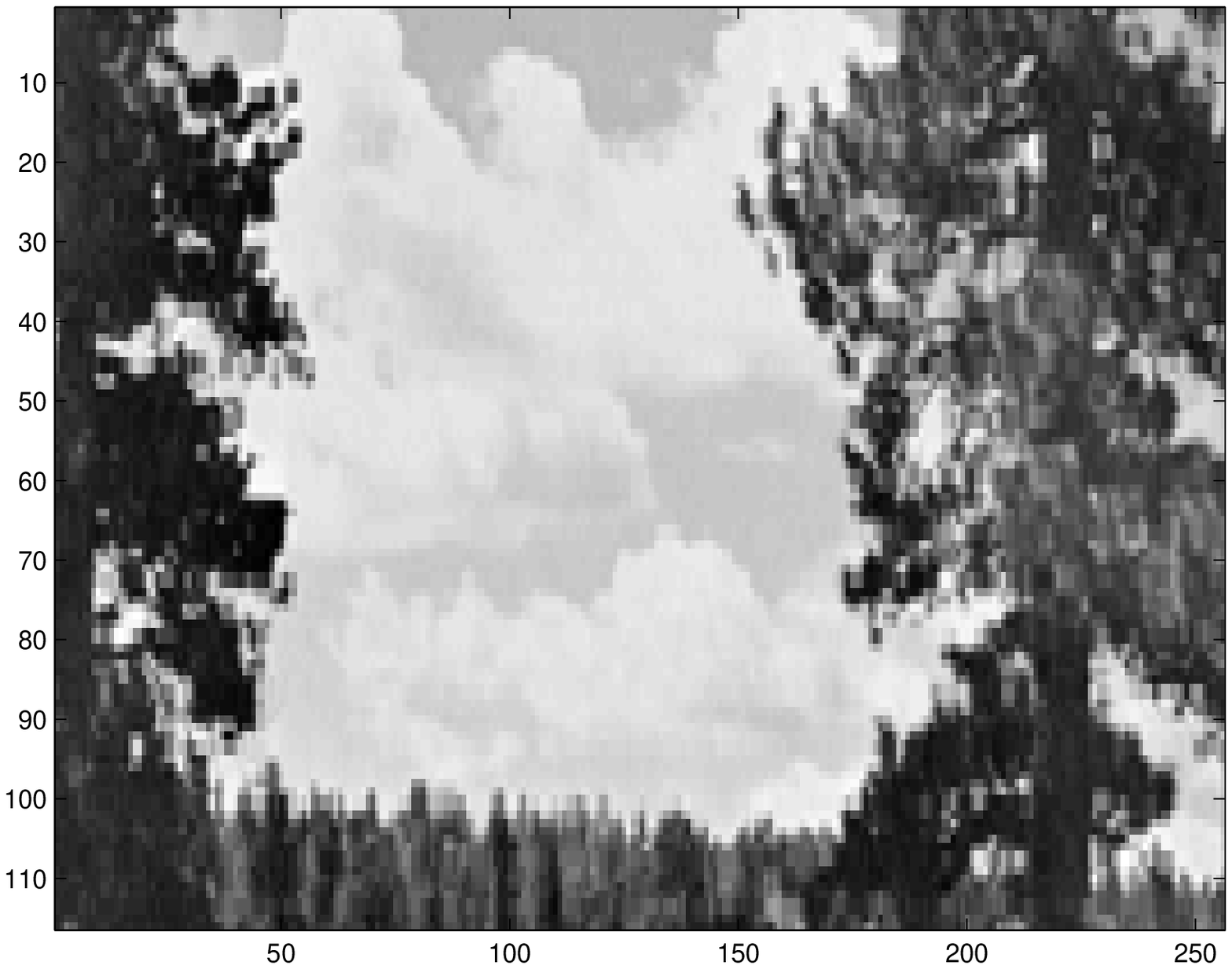} &  \includegraphics[width=8cm, height=3.7cm]{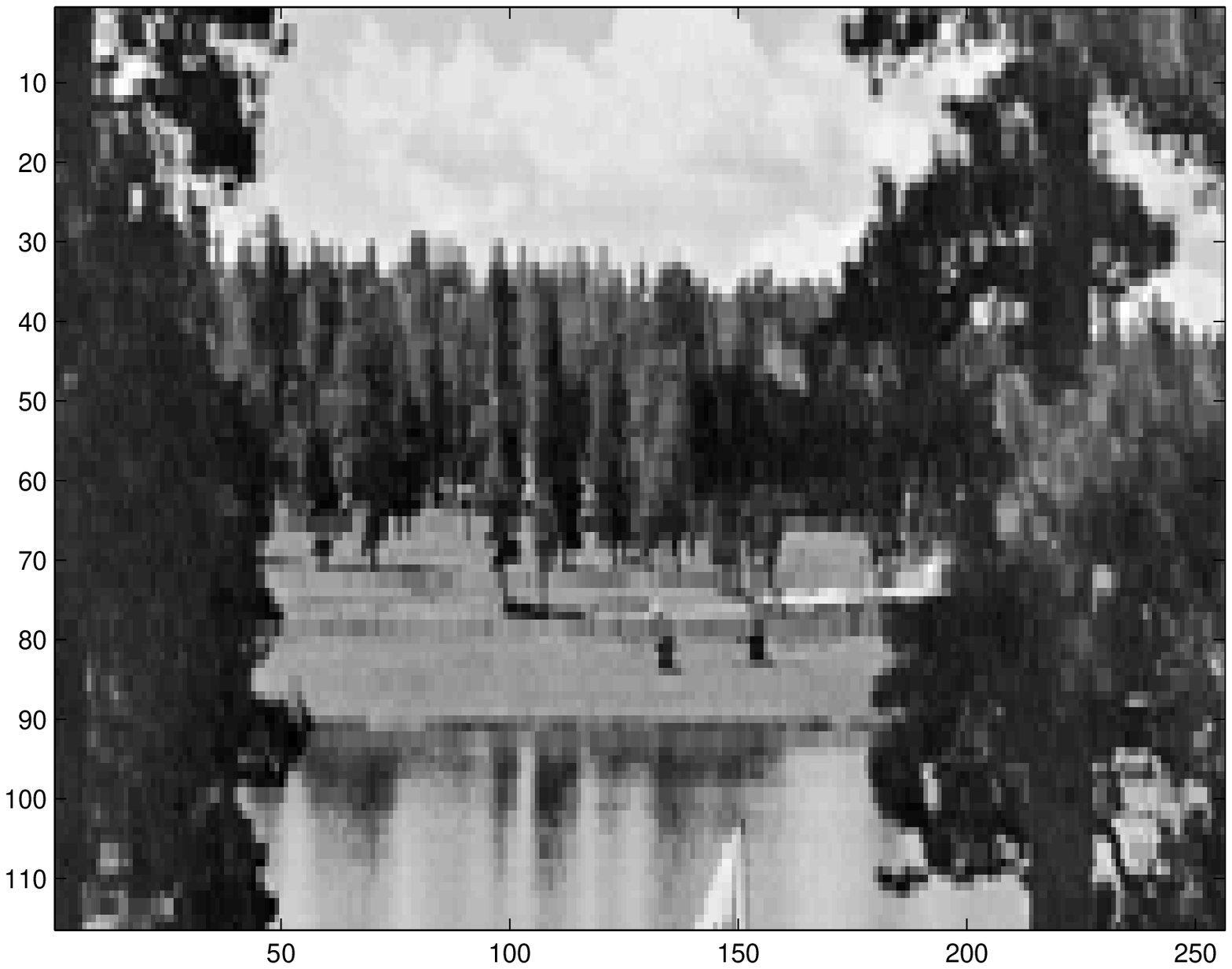}\\
   \includegraphics[width=8cm, height=3.7cm]{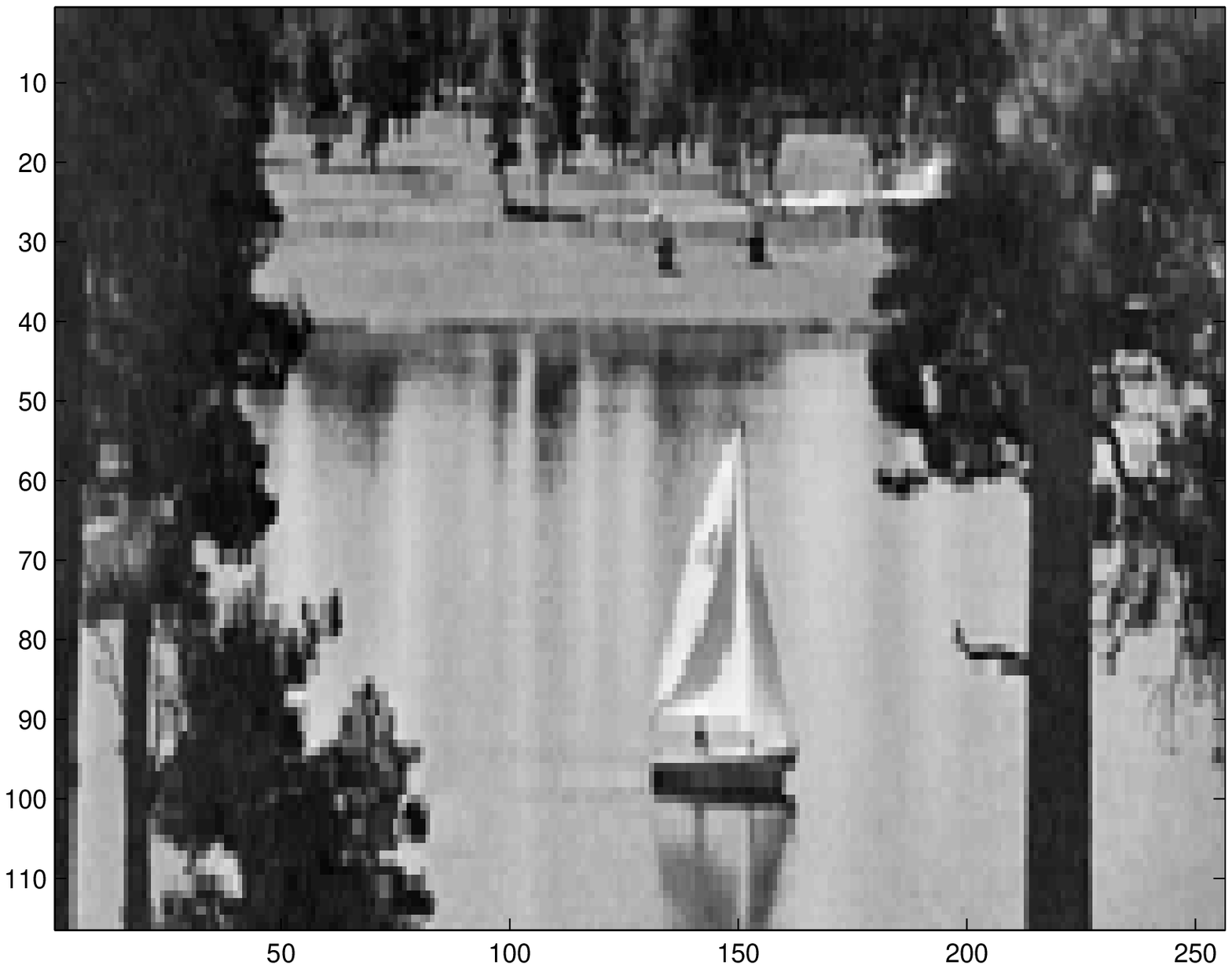} &  \includegraphics[width=8cm, height=3.7cm]{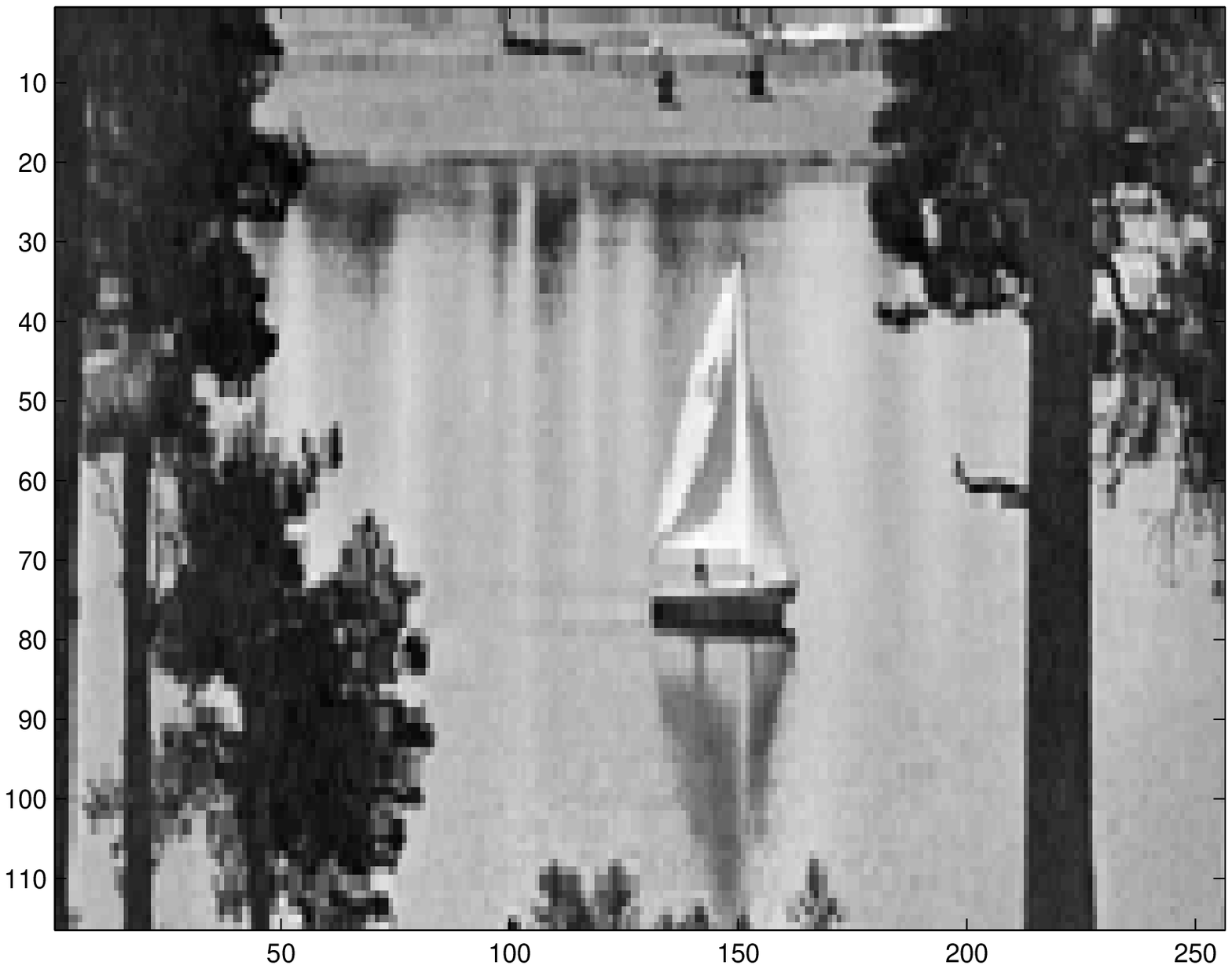}
\end{tabular}
 \vspace*{-5mm}\caption{Examples of selected signals  to be estimated from observed data considered in Example 5-9.}
 \label{fig4}
 \end{figure}

\begin{figure}[]
\centering
 \hspace*{-7mm}\begin{tabular}{c@{\hspace*{0mm}}c}
    \includegraphics[width=8cm, height=3.7cm]{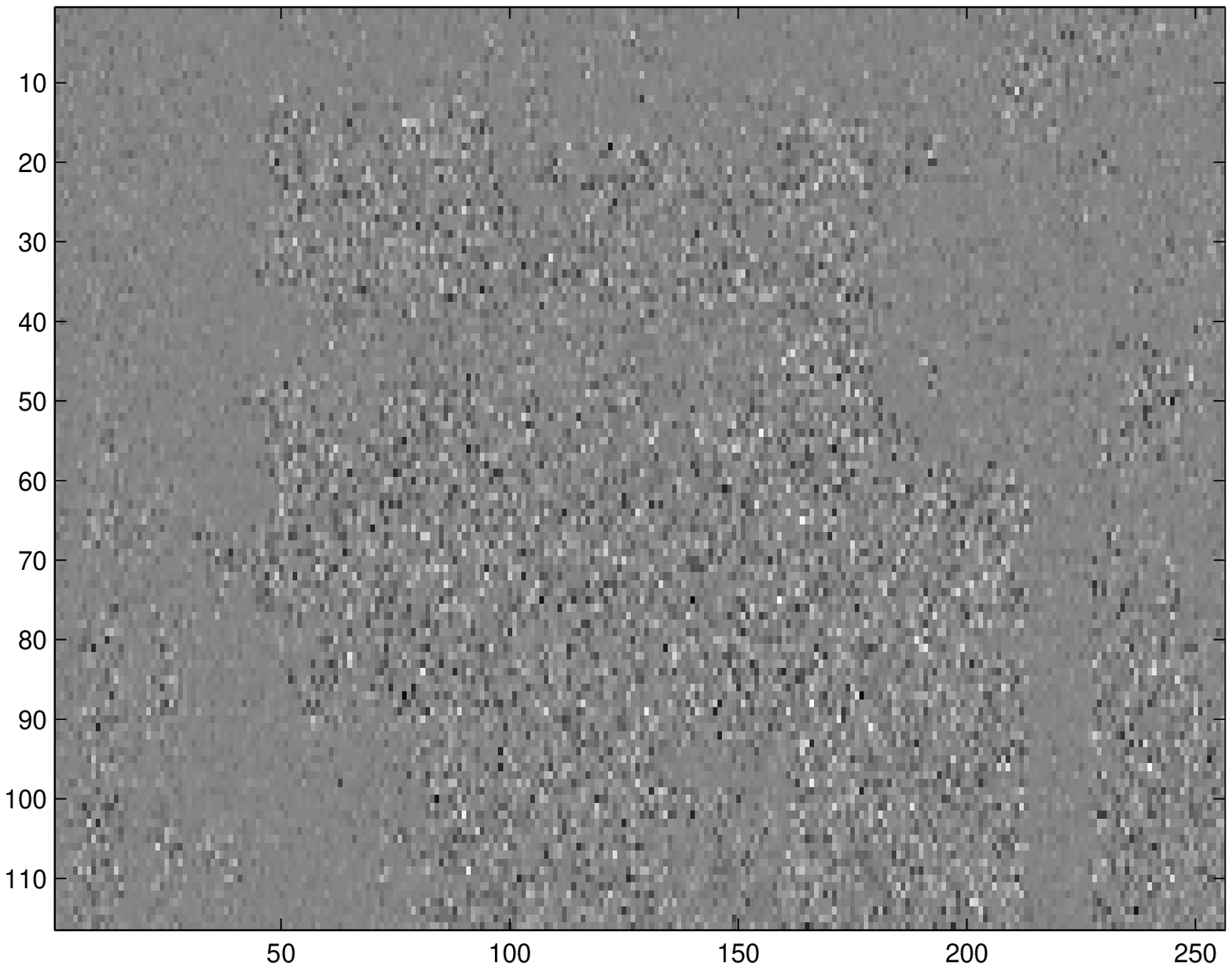} &  \includegraphics[width=8cm, height=3.7cm]{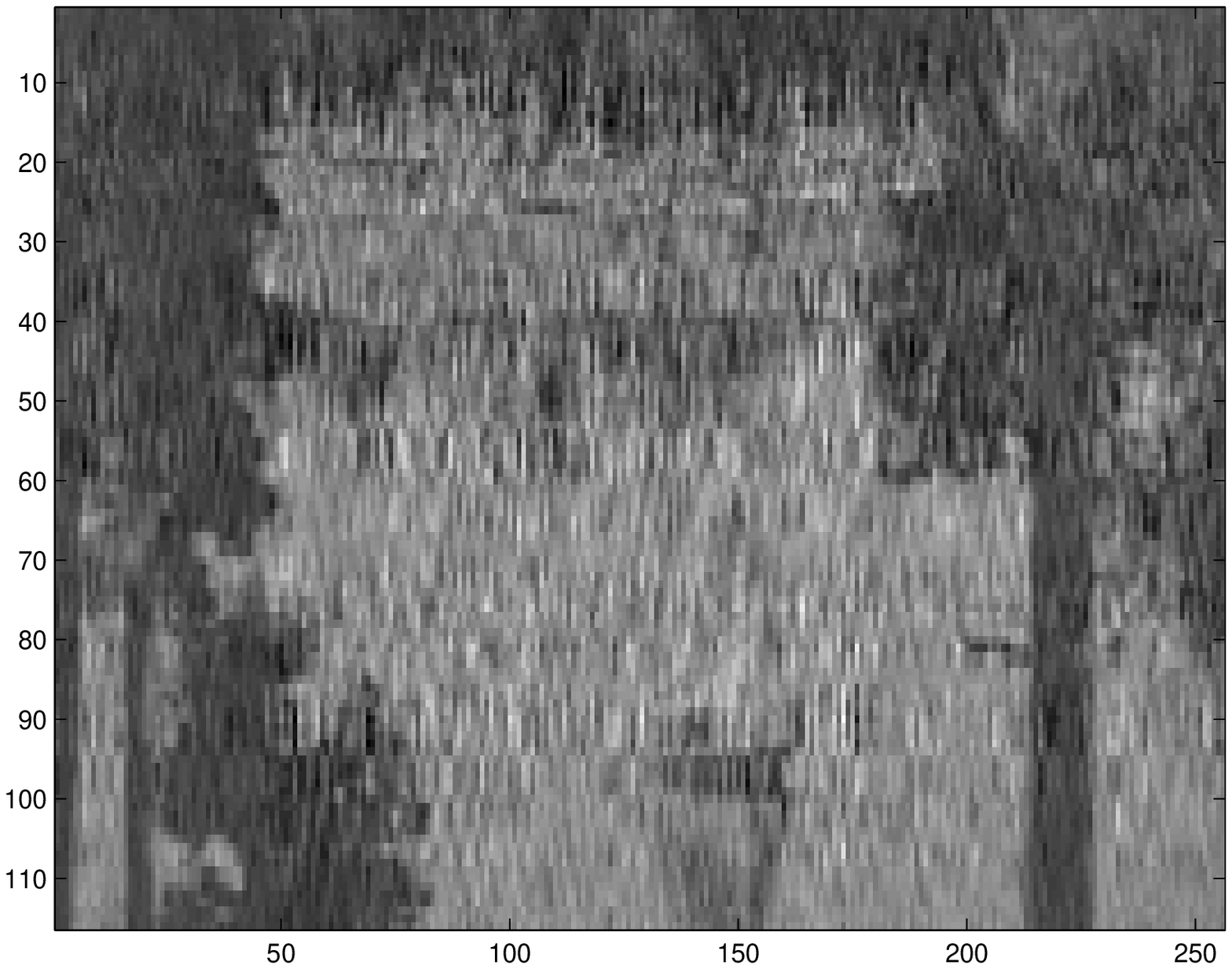}\\
    {\small   (a) Observed signal  $Y^{(120)}.$} &  {\small   (b) Estimate of $X^{(120)}$ by piecewise filter $F^{(28)}$.} \\
     \includegraphics[width=8cm, height=3.7cm]{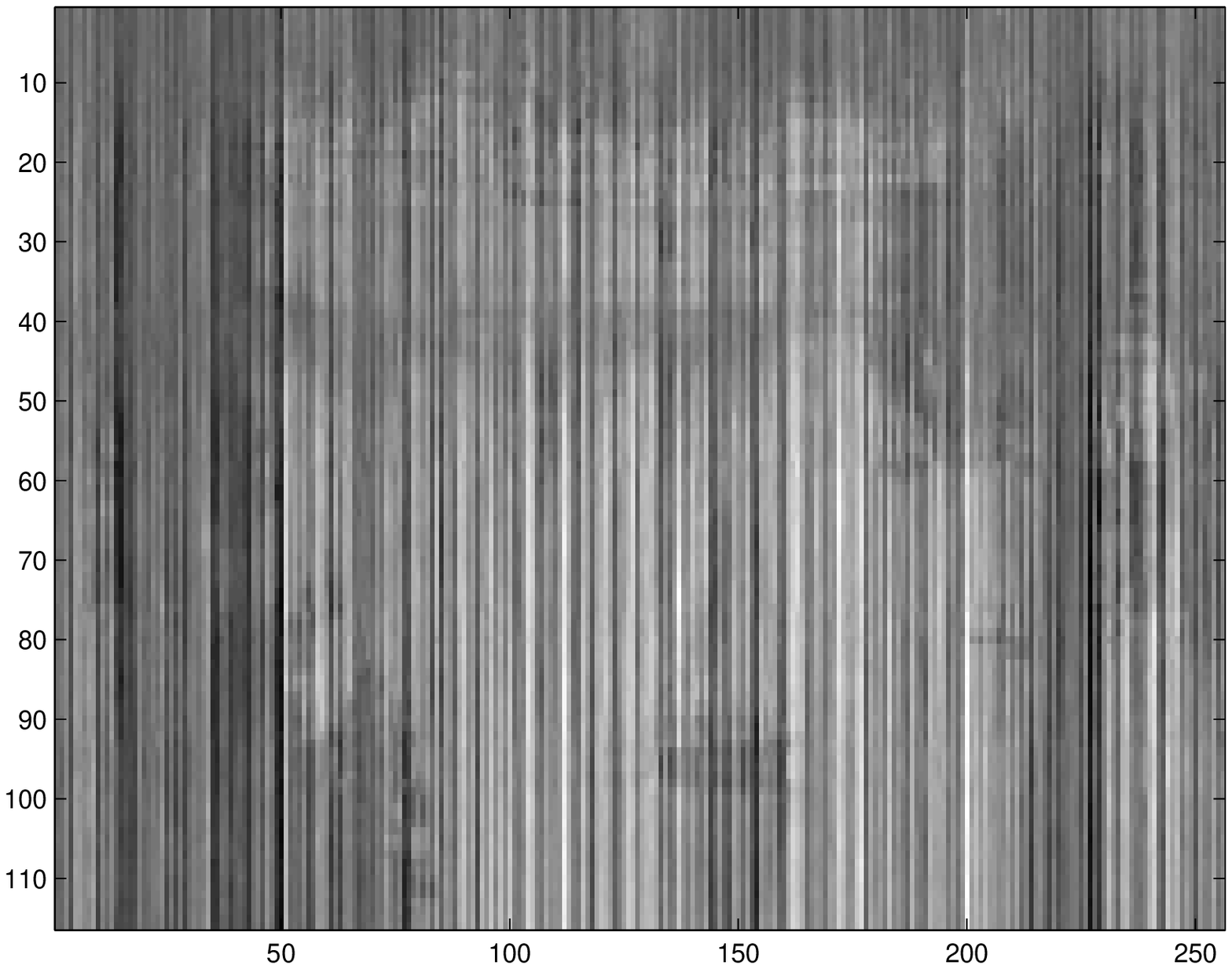} &  \includegraphics[width=8cm, height=3.7cm]{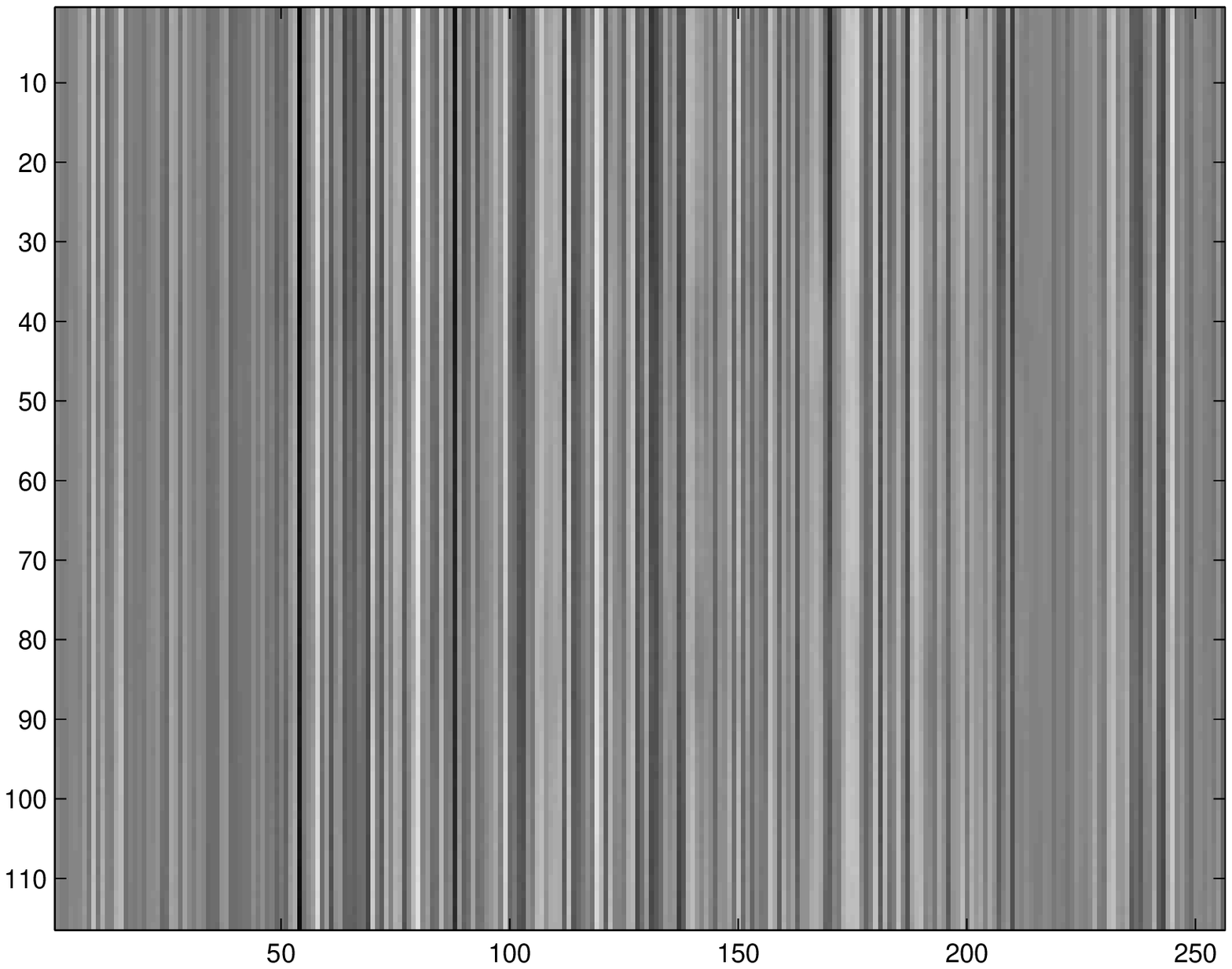}\\
    {\small   (c) Estimate of $X^{(120)}$ by GOL filter.} &  {\small   (d) Estimate of $X^{(120)}$ by averaging polynomial  filter.} 
\end{tabular}
 \vspace*{-5mm}\caption{Examples of the observed signal and the  estimates obtained by different filters.}
 \label{fig5}
 \end{figure}

\vspace*{-20mm}
\hspace*{-10mm}\begin{figure}[]
\centering
 \vspace*{-5mm}\begin{tabular}{c@{\hspace*{5mm}}c}
   \includegraphics[width=7.5cm, height=3.7cm]{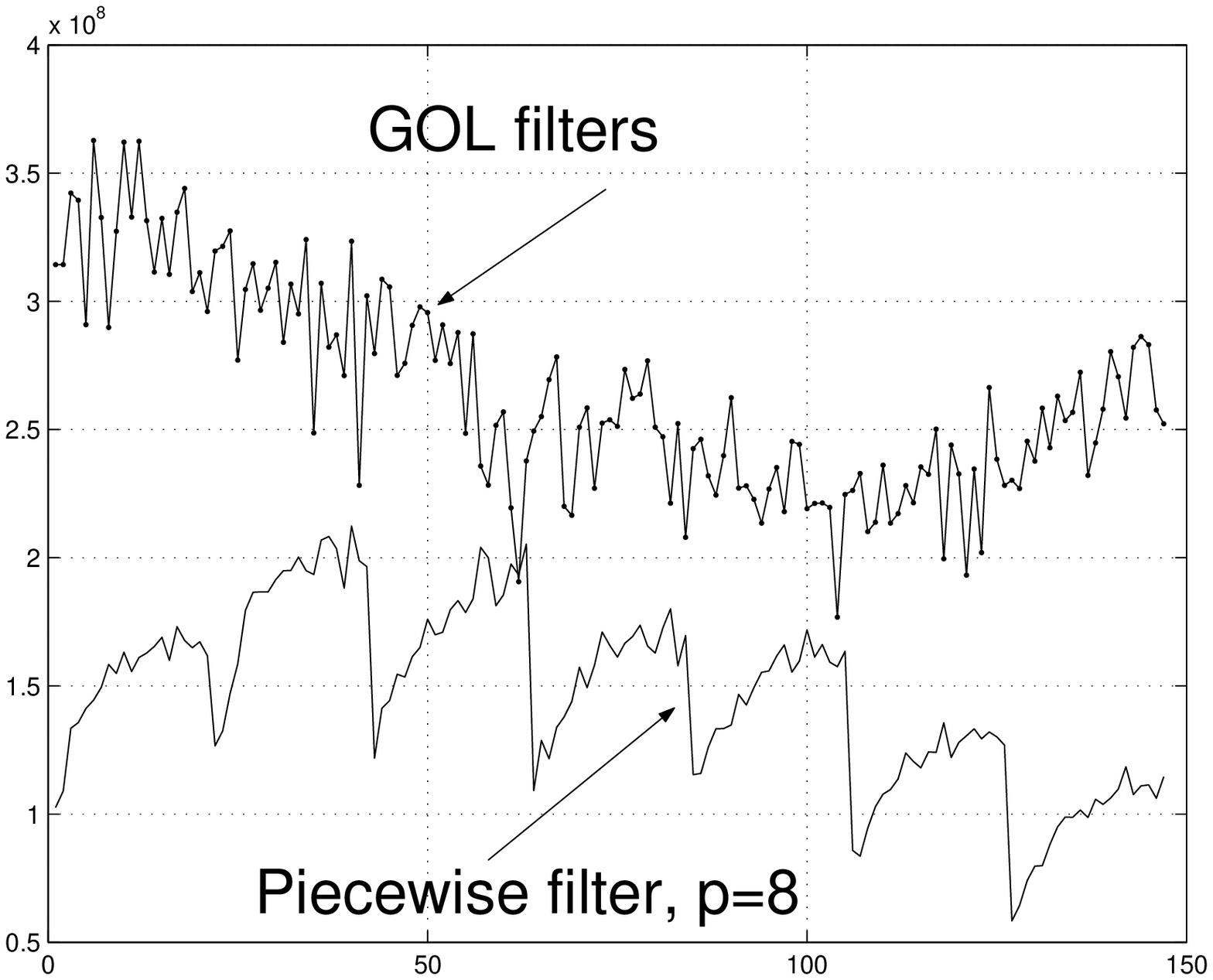} &  \includegraphics[width=7.5cm, height=3.7cm]{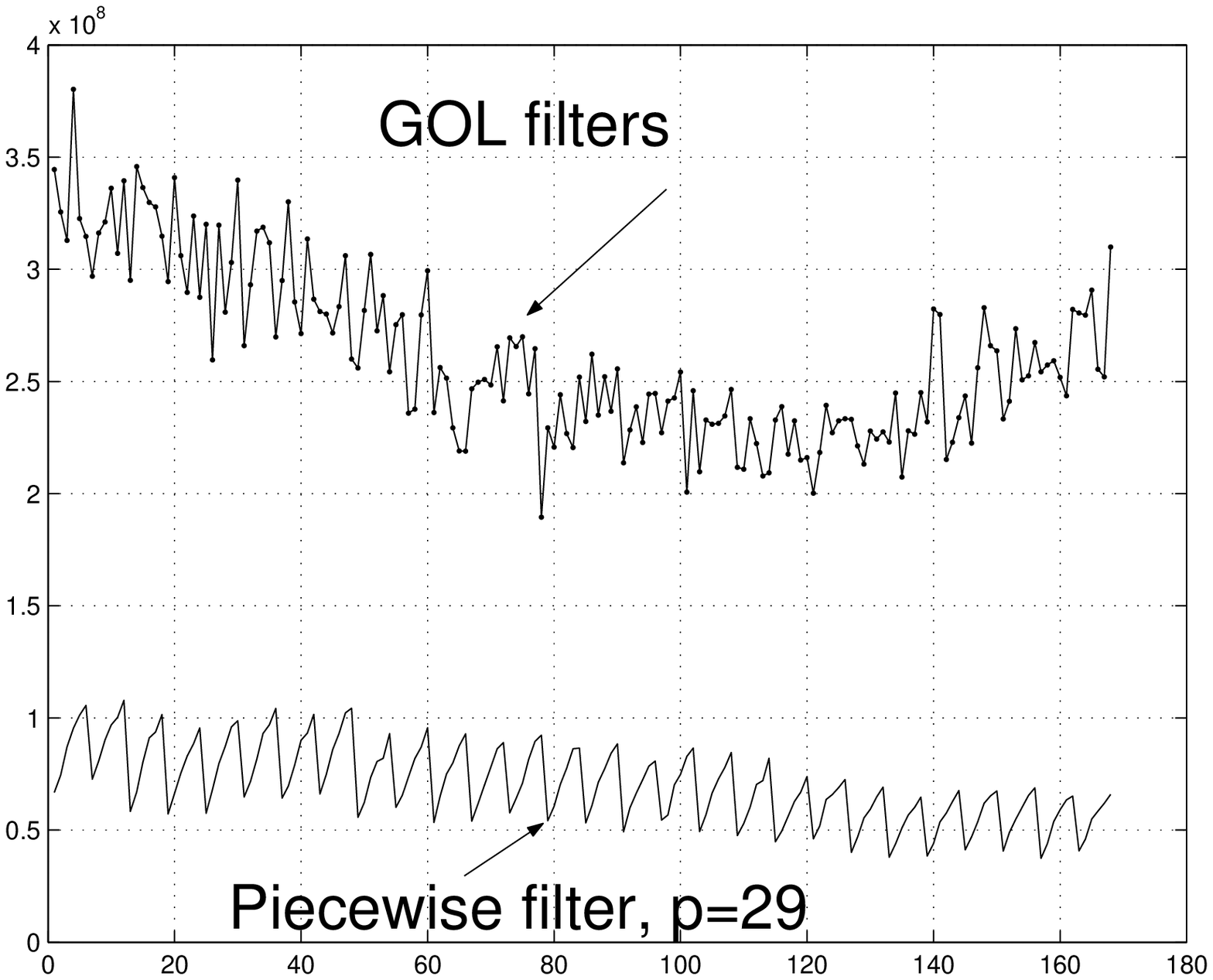}\\
  \includegraphics[width=7.5cm, height=3.7cm]{piecewise_boat_7.eps} &  \includegraphics[width=7.5cm, height=3.7cm]{piecewise_boat_28.eps}
\end{tabular}
 \vspace*{2mm}\caption{Illustration of errors associated with the piecewise interpolation filter $F^{(p-1)}$ of order $p$ and
  the generic optimal linear (GOL) filters \cite{tor100} applied to signals described in Examples 5--8.}
 \label{fig6}
 \end{figure}

\hspace*{-10mm}\begin{figure}[]
\centering
 \vspace*{-5mm}\begin{tabular}{c@{\hspace*{5mm}}c}
    \includegraphics[width=7.5cm, height=3.7cm]{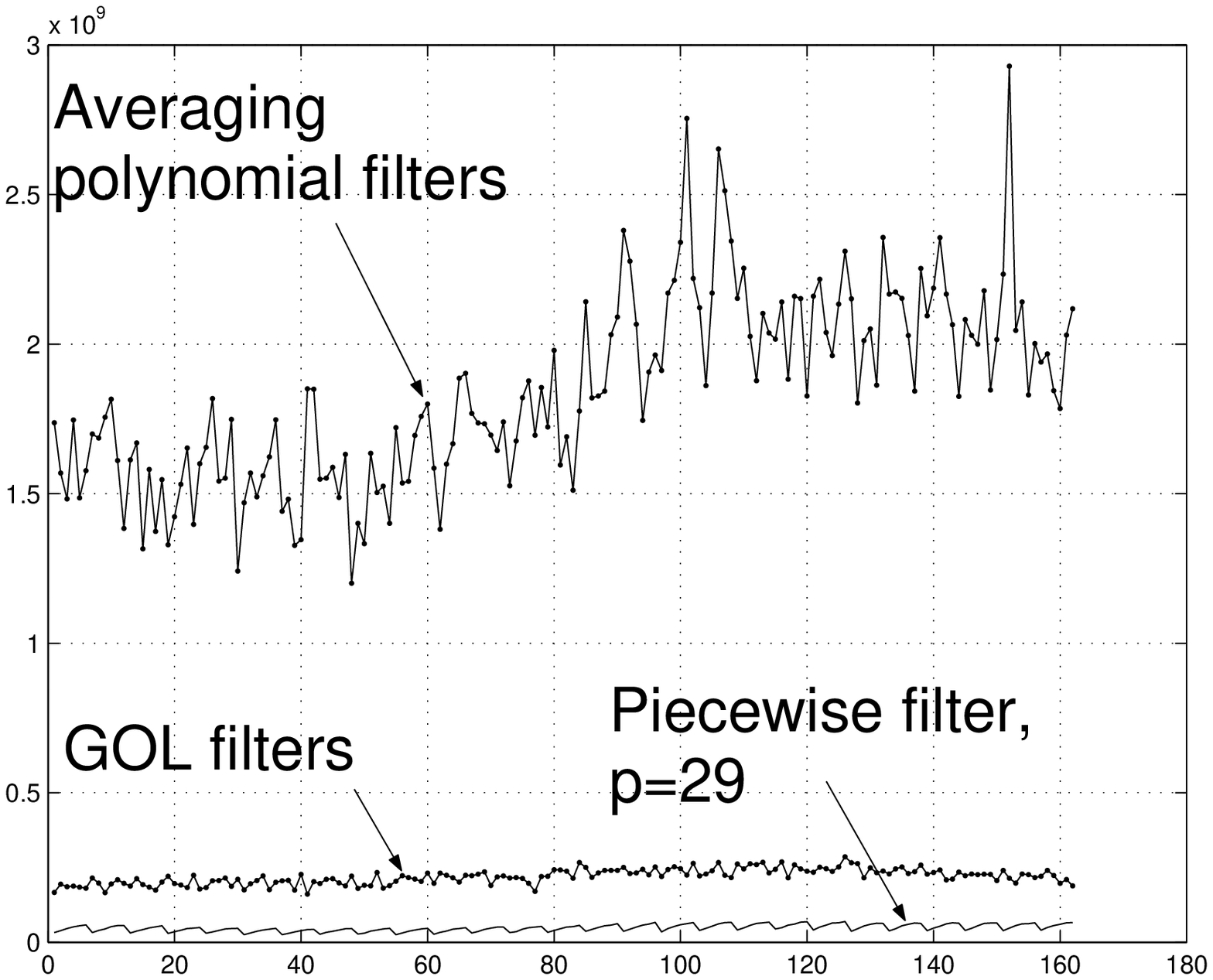} &  \includegraphics[width=7.5cm, height=3.7cm]{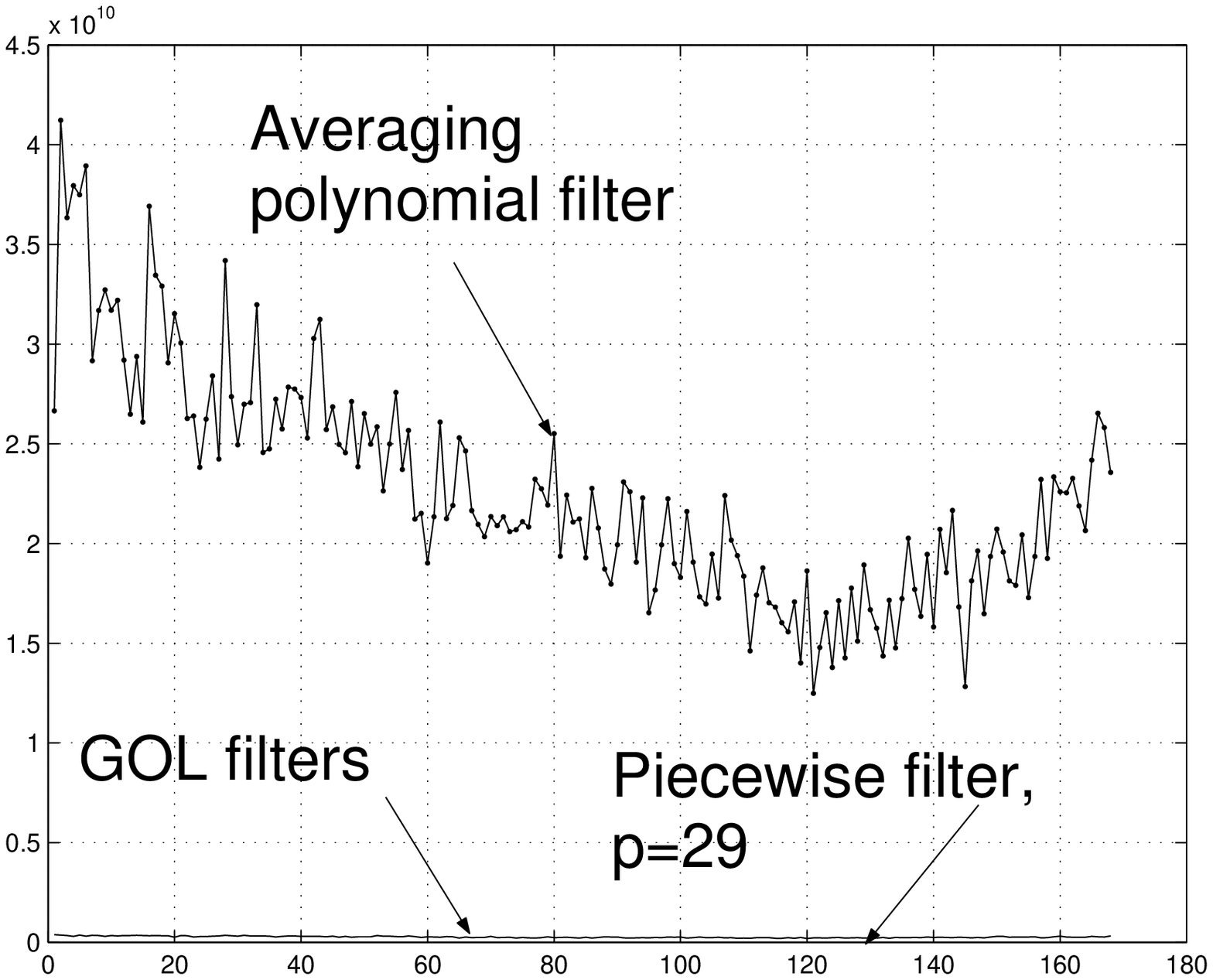}
\end{tabular}
 \vspace*{2mm}\caption{Illustration of errors associated with the averaging polynomial filters \cite{tor1,tor-m1}
in Examples 9 and 10.}
 \label{fig7}
 \end{figure}

\hspace*{-10mm}\begin{figure}[h]
\centering
 \vspace*{-5mm}\begin{tabular}{c@{\hspace*{5mm}}c}
     \includegraphics[width=7.5cm, height=3.7cm]{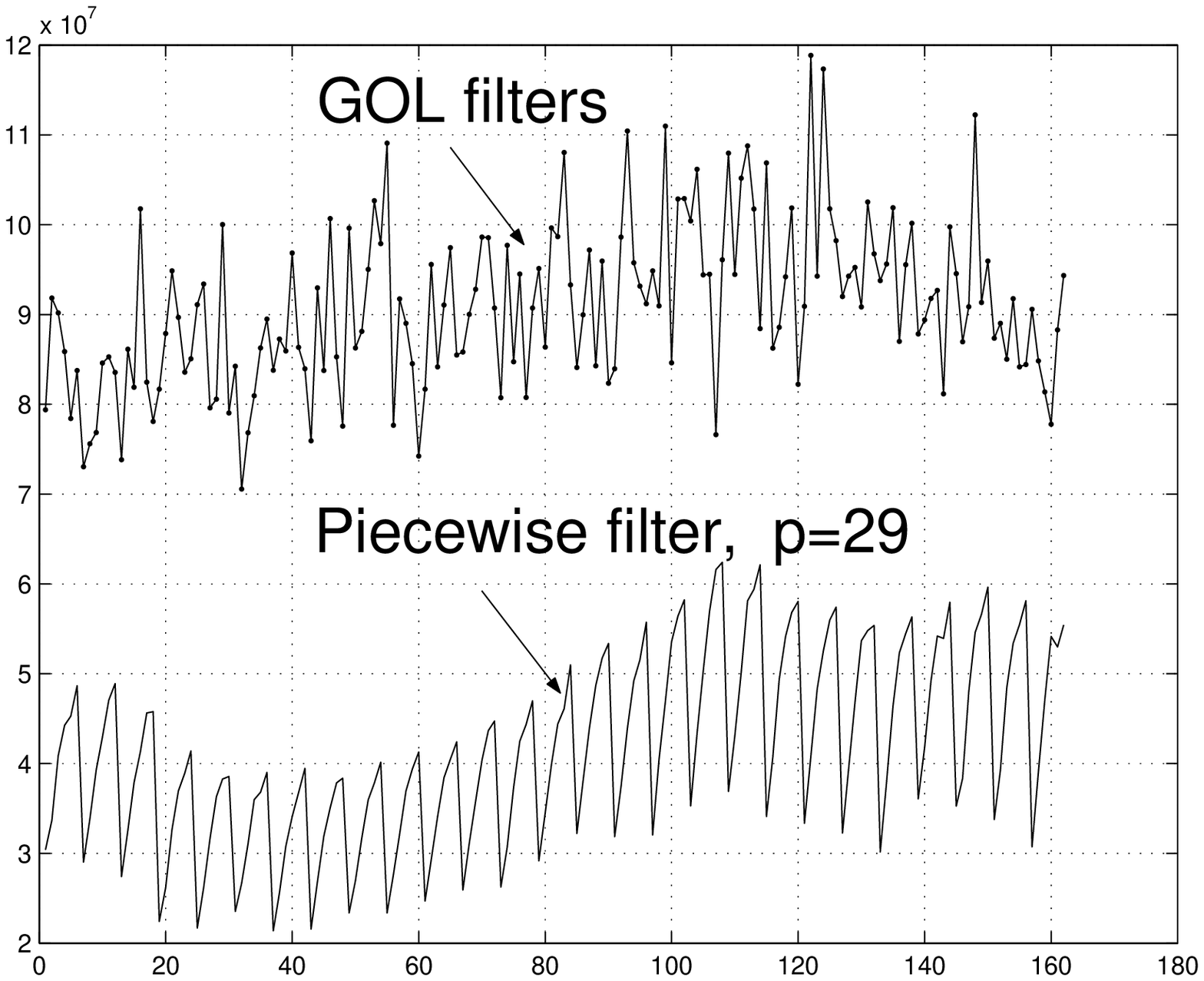} &  \includegraphics[width=7.5cm, height=3.7cm]{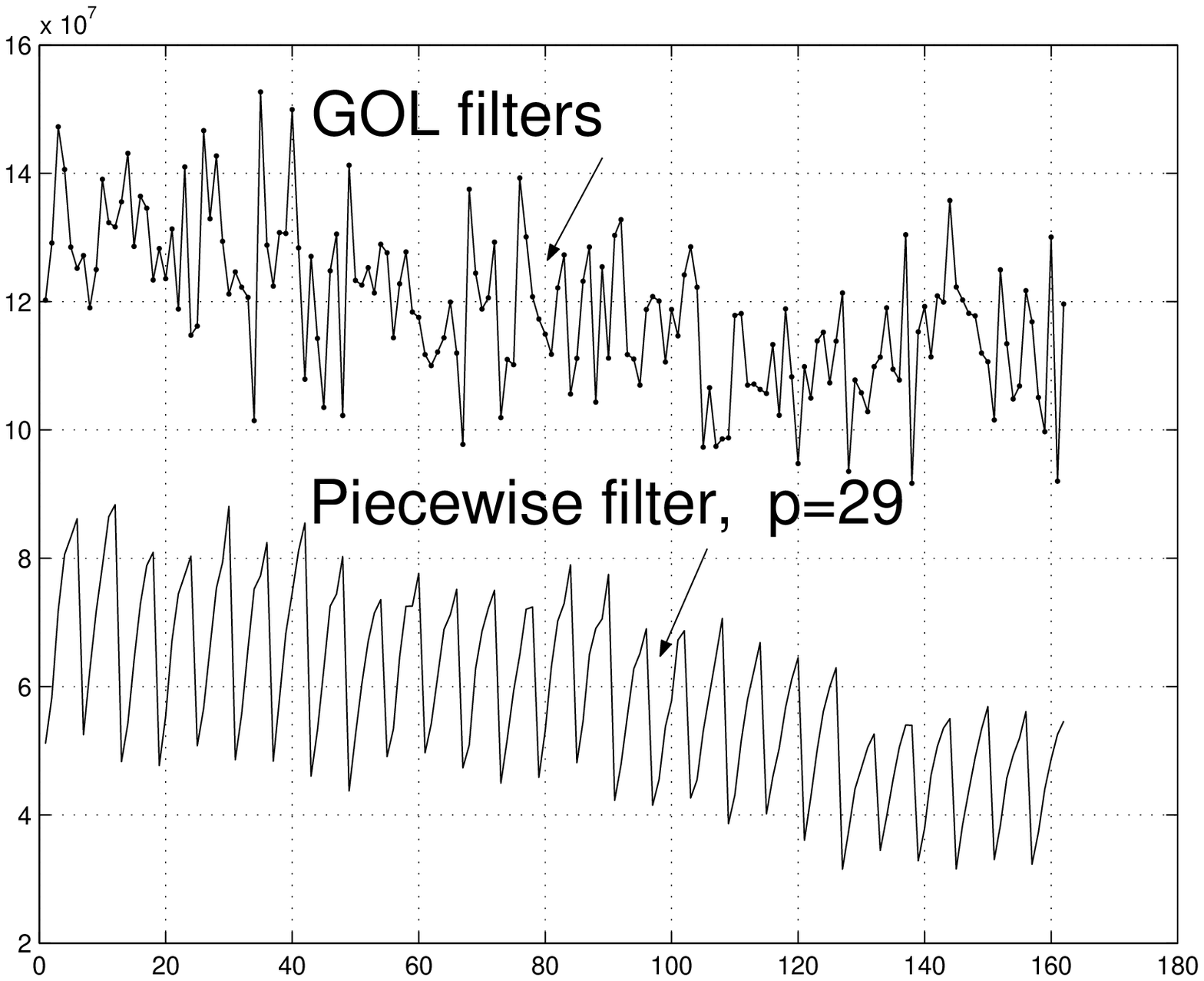}
\end{tabular}
 \vspace*{2mm}\caption{Illustration of errors associated with the piecewise interpolation filter $F^{(p-1)}$ and
 the generic optimal linear (GOL) filters \cite{tor100} applied to signals described in Examples 11 and 12.}
 \label{fig8}
 \end{figure}

\subsection{Further simulations with different type of  noise}\label{sim4}
In Examples 11 and 12 below, a different type of noise is considered. Unlike the multiplicative noise in (\ref{sim1yk}) and
(\ref{sim2yk}), here, the noise is additive.

{\em Example 11.} First, the piecewise interpolation filter $F^{(28)}$, the GOL filters \cite{tor100} and the averaging
polynomial filter \cite{tor-m1} have been applied to  the observed signals given by
\begin{equation}\label{sim4yk}
Y^{(k)} = X^{(k)}+ 900\times \mbox{\tt randn}_{(k)},\quad \mbox{for $k=1,\ldots,141$.}
\end{equation}
where $X^{(k)}$ is as in Section \ref{sim0}, i.e. $X^{(k)}$ is formed from the images 'plant'. In Fig. \ref{fig8} (a), the
diagrams of the errors associated with filter $F^{(28)}$ and the  GOL filters \cite{tor100} are given.
The errors associated with the   averaging polynomial filter \cite{tor-m1}, $\{\epsilon_{k_W}\}_{k=1}^{141}$, are much grater
(of order ${\mathcal O}(10^9)$) and they are not presented in  Fig. \ref{fig8} (a).

{\em Example 12.} In this example, the reference signals $X^{(1)},\ldots, X^{(141)}$ are as those in Section \ref{boat2},
i.e. they are formed from the image 'boat'. The observed signals are given by
\begin{equation}\label{sim4yk2}
Y^{(k)} = X^{(k)}+ 1000\times \mbox{\tt randn}_{(k)},\quad \mbox{for $k=1,\ldots,141$.}
\end{equation}
The  piecewise interpolation filter $F^{(28)}$ and the  GOL filters \cite{tor100} estimate the reference signals with the
associated errors represented in Fig.  \ref{fig8} (b).
As in Example 11 above, in this case, the errors associated with the   averaging polynomial filter \cite{tor1,tor-m1}  are much grater
(of order ${\mathcal O}(10^{10})$) and they are not presented in  Fig. \ref{fig8} (b).

Examples 11 and 12 further demonstrate the advantages of the proposed piecewise interpolation filter.

\subsection{Summary of simulations}\label{sum-sim}

The above simulations confirm the theoretical results obtained in Theorems \ref{th1}--\ref{th3}. In particular, Figs.
 \ref{fig3} and \ref{fig6} demonstrate that the error associated with the piecewise interpolation filter $F^{(p-1)}$ decreases
when the number of sub-filters $F_1,\ldots, F_p$, $p$, increases.

A comparison between the proposed filter $F^{(p-1)}$ and the  known related filters \cite{tor100,tor1,tor-m1,rus1,cou1}
has been done. The filter $F^{(p-1)}$  estimates the reference signals with the accuracies  that are much better than
those of the generic optimal linear (GOL) filters  \cite{tor100} and the averaging polynomial filter \cite{tor1,tor-m1}.
Further, the filters proposed  in \cite{rus1,cou1} fail in processing the signals under consideration. This is because the observed signals
in (\ref{sim1yk}),  (\ref{sim2yk}), (\ref{sim4yk}) and  (\ref{sim4yk2}) are grossly corrupted and, therefore, the inverse matrices
used in the filter structures in   \cite{cou1} do not exist. The technique in \cite{rus1} requires the use
of the reference signal in the proposed filter which is supposed to be unknown in the simulations above.

The filters have been applied to the different signal sets (presented in Sections \ref{sim0} and \ref{boat2}), using
different forms of noise (given in (\ref{sim1yk}),  (\ref{sim2yk}), (\ref{sim4yk}) and  (\ref{sim4yk2})).

The computational work associated with the proposed filter $F^{(p-1)}$ is substantially less than
that associated with the known filters  discussed in Section \ref{intr} (in particular, with the filters in
\cite{ves1}--\cite{cou1}).  This is because, for the processing of a data set containing $N$ signals, filter $F^{(p-1)}$ requires
computation of $p$ covariance matrices  with $p\ll N$
while the known  filters require  computation  of $N$ matrices (in the above Examples, $p=5,7,15,28$, respectively, and $N=141$).

\section{Conclusions}

The theory for a new approach to filtering  arbitrarily large sets of stochastic signals $K_{Y}$ and $K_{X}$  is provided.
Distinctive features  of the  approach are as follows.

(i) The proposed filter $\f^{(p-1)}:K_{Y}\rightarrow K_{X}$ is nonlinear and is presented  in the form of a sum with  $p-1$ terms  where each term,
$\f_j:K_{{Y,j}}\rightarrow K_{{X,j}}$, is interpreted as  a particular  sub-filter. Here, $K_{{Y,j}}$ and $K_{{X,j}}$ are `small'
pieces of $K_{Y}$ and $K_{X}$, respectively.

(ii) The prime idea is to
exploit a priori information only  on {\em  few reference signals}, $p$, from the set
$K_{_X}$ that contains $N\gg p$  signals (or even an infinite number of signals) and determine $\f_j$ separately, for each pieces
$K_{{Y,j}}$ and $K_{{X,j}}$, so that
 the associated  error is minimal. In other words, the filter $\f^{(p-1)}$ is flexible to changes in
the sets of observed and reference signals $K_{Y}$ and  $K_{X}$, respectively.

(iii) Due to the specific way of determining $\f_j$, the filter $\f^{(p-1)}$ provides a smaller associated error than that
for the processing of the  whole  set $K_{Y}$ by a filter which is not specifically adjusted to each particular piece $K_{Y,j}$.
Moreover, the error associated with our filter decreases when the number of its terms, $\f_1,\ldots, \f_{p-1}$, increases.

(iv) While the proposed filter $\f^{(p-1)}$ processes  arbitrarily large (and even infinite) signal sets, the filter is nevertheless
fixed for all signals in the sets.

(v) The filter $\f^{(p-1)}$ is determined in terms of pseudo-inverse matrices so that the filter always exists.

(vi) The computational load associated with the filter $\f^{(p-1)}$ is less than that associated with other known filters applied to
the processing of large signal sets.

\section{Appendix }\label{appa}

{\em Proof of Theorem \ref{th1}}: It follows from (\ref{xt1}) and (\ref{xtp1}) that
 $\alpha_j$, for $j=1,\ldots, p-1$,   is given by
\begin{eqnarray}\label{aj1}
\alpha_j = \widehat{\x}(t_{j},\omega)- B_{j}[\y(t_{j},\omega)].
\end{eqnarray}
Further, for $\alpha_j$ given by (\ref{aj1}),
\begin{eqnarray}
 &&\hspace*{-10mm}\left \|[\x(t_{j+1},\cdot)-\alpha_j]  - \bb_j [\y(t_{j+1},\cdot)]\right \|^2_\Omega\nonumber\\
& = &   \left \|\zz(t_j,t_{j+1},\cdot) - \bb_j [\ww(t_j,t_{j+1},\cdot))]\right \|^2_\Omega \label{zzjwj}\\
& = &   \mbox{tr} \{E_{z_{j} z_{j}} - E_{z_{j}w_{j}}B_j^T - B_j E_{w_{j}z_{j}} + B_j E_{w_{j}w_{j}} B_j^T\}\nonumber \\
\label{62-er2} & = &\|E_{z_{j}z_{j}}^{1/2}\|^2 -
\|E_{z_{j}w_{j}}(E_{w_{j}w_{j}}^{1/2})^\dag\|^2 + \|(B_j - E_{z_{j}w_{j}}E_{w_{j}w_{j}}^{\dag}) E_{w_{j}w_{j}}^{1/2}\|^2 \nonumber \\
& = & \|E_{z_{j}z_{j}}^{1/2}\|^2 - \|E_{z_{j}w_{j}}(E_{w_{j}w_{j}}^{1/2})^\dag\|^2 + \|E_{z_{j}w_{j}}(E_{w_{j}w_{j}}^{1/2})^{\dag}
- B_j E_{w_{j}w_{j}}^{1/2}\|^2\label{drx1},
\end{eqnarray}
where $\|\cdot\|$ is the Frobenius norm. The latter is true because
$$
E_{w_{j}w_{j}}^{\dag} E_{w_{j}w_{j}}^{1/2} = (E_{w_{j}w_{j}}^{1/2})^{\dag}
$$
and
\begin{equation}\label{62-ee}
E_{z_{j}w_{j}}E_{w_{j}w_{j}}^\dag E_{w_{j}w_{j}} = E_{z_{j}w_{j}}
\end{equation}
by Lemma 24 in \cite{tor100}.  Thus, the second expression in (\ref{xtp1}) is reduced to the problem
\begin{equation}\label{min12}
\min_{B_j} \|E_{z_{j}w_{j}}(E_{w_{j}w_{j}}^{1/2})^{\dag} - B_jE_{w_{j}w_{j}}^{1/2}\|^2.
\end{equation}
It is known (see, for example, \cite{tor100}, p. 304) that the solution of  problem (\ref{min12}) is given by (\ref{b0j}).
The equation (\ref{f0jxj}) follows from (\ref{fyab}) and (\ref{aj1}).

Theorem \ref{th1} is proven. $\hfill \Box$

\bigbreak
{\em Proof of Theorem \ref{th2}}: For $t\in [t_j, \hspace*{1mm} t_{j+1}]$ and $F_j$ defined by (\ref{f0jxj})--(\ref{b0j}),
\begin{eqnarray}
&&\x(t,\omega) - {F}[\y(t,\omega)]\label{xtfjy}\nonumber\\
&& = \x(t,\omega) - {F}_j[\y(t,\omega)]\label{xtfjy}\nonumber\\
&& = \x(t,\omega) -  \widehat{\x}(t_{j},\omega) + B_{j}\y(t_{j},\omega) - {B}_j\y(t,\omega)\nonumber\\
&& = [\x(t,\omega) - \x(t_{j+1},{\omega})] + \zz(t_j,t_{j+1},{\omega})  - {B}_j\ww(t_j,t_{j+1},\omega)
+ {B}_j[\y(t_{j+1},\omega)-\y(t,\omega)].\label{xttjfjy}
\end{eqnarray}
Then (\ref{xtfjy}) and  (\ref{xttjfjy}) imply
\begin{eqnarray}\label{xtf}
\| \x(t,\omega) - {F}[\y(t,\omega)]\|^2_{T,\Omega} &\leq& \|\x(t,\omega) - \x(t_{j+1},{\omega}) \|^2_{T,\Omega}\nonumber\\
&+ & \|\zz(t_j,t_{j+1},{\omega})  - {B}_j\ww(t_j,t_{j+1},\omega) \|^2_{\Omega}\\
&+ & \|{B}_j[\y(t_{j+1},\omega)-\y(t,\omega)]\|^2_{T,\Omega}\nonumber
\end{eqnarray}
where $\|\zz(t_j,t_{j+1},{\omega})  - {B}_j\ww(t_j,t_{j+1},\omega) \|^2_{\Omega}=\|\zz(t_j,t_{j+1},{\omega})
- {B}_j\ww(t_j,t_{j+1},\omega) \|^2_{T,\Omega}$.

It follows from (\ref{zzjwj}) and (\ref{drx1}) that for $B_j$ given by (\ref{b0j}),
\begin{eqnarray}\label{zztj}
\|\zz(t_j,t_{j+1},{\omega})  - {B}_j\ww(t_j,t_{j+1},\omega) \|^2_{\Omega}
=\|E_{z_{j}z_{j}}^{1/2}\|^2 - \|E_{z_{j}w_{j}}(E_{w_{j}w_{j}}^{1/2})^\dag\|^2.
\end{eqnarray}

Then (\ref{fyd0})--(\ref{b0j}), (\ref{lip11}) and (\ref{xtfjy})--(\ref{zztj}) imply that for all $t\in [a, \hspace*{1mm} b]$ and $\omega\in \Omega$,
(\ref{er13}) is true. $\hfill \Box$

\bigbreak

{\em Proof of Theorem \ref{th3}}: The relation (\ref{norm}) implies that
\begin{eqnarray}\label{fff1}
\| \x(t,\omega) - {F}[\y(t,\omega)]\|^2_{T,\Omega}=\frac{1}{b - a} \sum_{j=1}^{p-1} \int_{t_j}^{t_{j+1}}\| \x(t,\omega)
- F_j[\y(t,\omega)]\|^2_{\Omega} dt,
\end{eqnarray}
where
\begin{eqnarray*}\label{}
&&\hspace*{-10mm}\|\x(t,\omega) - F_j[\y(t,\omega)]\|^2_{\Omega}\\
&&=\|\x(t,\omega) -  \widehat{\x}(t_{j},\omega) + B_{j}[\y(t_{j},\omega)
- {B}_j\y(t,\omega)]\|^2_{\Omega}\\
&&\hspace*{10mm} \leq  \|\x(t,\omega) -  \x(t_{j},\omega)\|^2_{\Omega} + \|\x(t_j,\omega) -  \widehat{\x}(t_{j},\omega)\|^2_{\Omega}\\
 &&\hspace*{70mm}+ \|B_{j}[\y(t_{j},\omega) - {B}_j\y(t,\omega)]\|^2_{\Omega}.
\end{eqnarray*}
Then
\begin{eqnarray}\label{}
&&\hspace*{-10mm}\int_{t_j}^{t_{j+1}}\| \x(t,\omega)- F_j[\y(t,\omega)]\|^2_{\Omega} dt\label{int1}\\
&&\leq \int_{t_j}^{t_{j+1}}\| \x(t,\omega)-  \x(t_{j},\omega)\|^2_{\Omega} dt + \int_{t_j}^{t_{j+1}}\| \x(t_j,\omega)
-  \widehat{\x}(t_{j},\omega)\|^2_{\Omega} dt\nonumber \\
&&\hspace*{70mm}+\|B_j\| \int_{t_j}^{t_{j+1}}\|\y(t_j,\omega)-\y(t,\omega)\|^2_{\Omega} dt\nonumber \\
&&\leq \lambda_j  (\dr t_j)^2 + \| \x(t_j,\omega) - \widehat{\x}(t_{j},\omega)\|^2_{\Omega}\dr t_j +\|B_j\| \gamma_j (\dr t_j)^2\label{int2}
\end{eqnarray}

Let us consider an estimate of $\| \x(t_j,\omega) - \widehat{\x}(t_{j},\omega)\|^2_{\Omega}$, for $j=1,\ldots,p-1$.
To this end, let us denote $\displaystyle \dr t=\max_{j=1,\ldots,p-1} \dr t_j$.

For $j=1,$ i.e. for $t\in [t_1,\hspace*{1mm} t_2]$,
\begin{eqnarray*}\label{}
&&\| \x(t,\omega) - F_1{\y}(t,\omega)\|^2_{\Omega}\\
&&\leq \| \x(t,\omega) - {\x}(t_{1},\omega)\|^2_{\Omega}
+ \| \x(t_1,\omega) - \widehat{\x}(t_{1},\omega)\|^2_{\Omega}
+ \|B_1\| \|\y(t_1,\omega) - {\y}(t,\omega)\|^2_{\Omega}\\
&&\leq \lambda_1 \dr t_1 + c_1\dr t_1 +\|B_1\|\gamma_1 \dr t_1\\
&& \leq \beta_1 \dr t,
\end{eqnarray*}
where $\beta_1 = \lambda_1 + c_1+\|B_1\|\gamma_1$. In particular, the latter implies
 \begin{eqnarray*}\label{}
\| \x(t_2,\omega) - \widehat{\x}(t_{2},\omega)\|^2_{\Omega}=\| \x(t_2,\omega) - F_1{\y}(t_2,\omega)\|^2_{\Omega} \leq \beta_1 \dr t
\end{eqnarray*}

For $j=2,$ i.e. for $t\in [t_2,\hspace*{1mm} t_3]$,
\begin{eqnarray*}\label{}
&&\| \x(t,\omega) - F_2{\y}(t,\omega)\|^2_{\Omega}\\
&&\leq \| \x(t,\omega) - {\x}(t_{2},\omega)\|^2_{\Omega}
+ \| \x(t_2,\omega) - \widehat{\x}(t_{2},\omega)\|^2_{\Omega}
+ \|B_2\| \|\y(t_2,\omega) - {\y}(t,\omega)\|^2_{\Omega}\\
&&\leq \lambda_2 \dr t_2 + \beta_1\dr t +\|B_2\|\gamma_2 \dr t_2\\
&& \leq \beta_2 \dr t,
\end{eqnarray*}
where $\beta_2 = \lambda_2 + \beta_1+\|B_2\|\gamma_2$. In particular, then it follows that
 \begin{eqnarray*}\label{}
 \| \x(t_3,\omega) - \widehat{\x}(t_{3},\omega)\|^2_{\Omega}=\| \x(t_3,\omega) - F_2{\y}(t_3,\omega)\|^2_{\Omega}
\leq \beta_2 \dr t.
\end{eqnarray*}
On the basis of the above, let us assume that, for $j=k-1$ with $k=2,\ldots,p-1$, i.e. for $t\in [t_{k-1},\hspace*{1mm} t_k]$,
 \begin{eqnarray*}\label{}
\| \x(t_k,\omega) - \widehat{\x}(t_{k},\omega)\|^2_{\Omega}=\| \x(t_k,\omega) - F_{k-1}{\y}(t_k,\omega)\|^2_{\Omega}
\leq \beta_{k-1} \dr t
\end{eqnarray*}
where $\beta_{k-1}$ is defined by analogy with $\beta_2$.

Then, for $j=k$ with $k=2,\ldots,p-1$,  i.e. for $t\in [t_{k},\hspace*{1mm} t_{k+1}]$,
\begin{eqnarray*}\label{}
&&\| \x(t,\omega) - F_k{\y}(t,\omega)\|^2_{\Omega}\\
&&\leq \| \x(t,\omega) - {\x}(t_{k},\omega)\|^2_{\Omega}
+ \| \x(t_k,\omega) - \widehat{\x}(t_{k},\omega)\|^2_{\Omega}
+ \|B_k\| \|\y(t_k,\omega) - {\y}(t,\omega)\|^2_{\Omega}\\
&&\leq \lambda_k \dr t_k + \beta_{k-1}\dr t +\|B_k\|\gamma_2 \dr t_k\\
&& \leq \beta_k \dr t,
\end{eqnarray*}
where $\beta_k = \lambda_k + \beta_{k-1}+\|B_k\|\gamma_k$. Thus, the following is true:
 \begin{eqnarray}\label{int3}
\| \x(t_{k+1},\omega) - \widehat{\x}(t_{k+1},\omega)\|^2_{\Omega}=\| \x(t_{k+1},\omega) - F_{k}{\y}(t_{k+1},\omega)\|^2_{\Omega}
\leq \beta_{k} \dr t.
\end{eqnarray}
Therefore, (\ref{int1}), (\ref{int2}) and (\ref{int3}) imply
\begin{eqnarray}\label{}
&&\int_{t_j}^{t_{j+1}}\| \x(t,\omega)- F_j[\y(t,\omega)]\|^2_{\Omega} dt\nonumber\\
&&\leq \lambda_j  (\dr t_j)^2 + \beta_{j-1} (\dr t_j)^2 +\|B_j\| \gamma_j (\dr t_j)^2\label{int4}\\
&&\leq \eta_j  (\dr t)^2\nonumber
\end{eqnarray}
where $\eta_j = \lambda_j + \beta_{j-1} +\|B_j\|$, and then it follows from  (\ref{fff1})--(\ref{int2}) and (\ref{int4}) that
for all $t\in [a, \hspace*{1mm} b]$,
\begin{eqnarray}\label{fff2}
\| \x(t,\omega) - {F}[\y(t,\omega)]\|^2_{T,\Omega}\leq\frac{1}{b - a} \sum_{j=1}^{p-1} \eta_j  (\dr t)^2
=\frac{1}{b - a} \dr t\sum_{j=1}^{p-1}\eta_j \dr t.
\end{eqnarray}
Let us now choose $c\in \rt$ and $d\in \rt$ so that $\displaystyle\dr t= \frac{d - c}{p}$ and partition interval $[c, \hspace*{2mm}
d]\subset \rt$ by points $\tau_1,\ldots,\tau_p$ so that $c=\tau_1$ and $\tau_{j} = \tau_1 + j \dr t$ with $j=1,\ldots,p$.
There exists an integrable (bounded) function $\varphi: [c, \hspace*{2mm} d]\rightarrow \rt$ such that, for $\xi_j \in
(\tau_j,\hspace*{1mm} \tau_{j+1})$, $\varphi(\xi_j) = \eta_j$. Then
\begin{eqnarray}\label{lim11}
\lim_{\dr t\rightarrow \infty} \sum_{j=1}^{p-1}\eta_j \dr t = \lim_{\dr t\rightarrow \infty} \sum_{j=1}^{p-1}\varphi(\xi_j) \dr t
=\int_{c}^d \varphi (\tau) d \tau <+\infty.
\end{eqnarray}

Thus,
\begin{eqnarray}\label{fr11}
\frac{1}{b - a} \dr t\sum_{j=1}^{p-1}\eta_j \dr t \rightarrow 0
\quad\mbox{as}\hspace*{2mm}  \dr t\rightarrow 0.
\end{eqnarray}
As a result,  (\ref{fff2})--(\ref{fr11}) imply (\ref{er17}).  $\hfill \Box$

\end{document}